\documentclass[aps, pra, reprint, longbibliography,
superscriptaddress]{revtex4-1}

\usepackage{hyperref}
\usepackage{graphicx}
\usepackage[utf8]{inputenc}
\usepackage{xcolor}
\usepackage{amsmath, amssymb}
\usepackage{physics}
\usepackage{ulem}

\begin{document}

\preprint{APS/123-QED}

\title{Superfluid-supersolid phase transition of elongated dipolar Bose-Einstein Condensates at finite temperatures}

\author{J. S\'anchez-Baena}
\email{juan.sanchez.baena@upc.edu}
\affiliation{Departament de F\'isica, Universitat Polit\`ecnica de Catalunya, Campus Nord B4-B5, 08034 Barcelona, Spain}
\affiliation{Center for Complex Quantum Systems, Department of Physics and Astronomy, Aarhus University, DK-8000 Aarhus C,  Denmark}
\author{T. Pohl}
\affiliation{Center for Complex Quantum Systems, Department of Physics and Astronomy, Aarhus University, DK-8000 Aarhus C,  Denmark}
\affiliation{Institute for Theoretical Physics, Vienna University of Technology, Wiedner Hauptstraße 8-10, 1040 Vienna, Austria}
\author{F. Maucher}
\affiliation{Faculty of Mechanical Engineering; Department of Precision and Microsystems Engineering, Delft University of Technology, 2628 CD, Delft, The Netherlands}
\affiliation{Departament de F\'isica, Universitat de les Illes Balears \& IAC-3, Campus UIB, E-07122 Palma de Mallorca, Spain}

\date{\today}

\begin{abstract}

We analyse the finite-temperature phase diagram of a dipolar Bose Einstein Condensate confined in a tubular geometry.
The effect of thermal fluctuations is accounted for by means of Bogoliubov theory employing the local density approximation. In the considered geometry, the superfluid-supersolid phase transition can be of first- and second-order. We discuss how the corresponding transition point is affected by the finite temperature of the system.
\end{abstract}

\maketitle

\section{\label{sec:introduction}Introduction}

Supersolidity refers to a state of matter that simultaneously features both discrete translational symmetry whilst exhibiting a large superfluid fraction and was conceived fifty years ago~\cite{Andreev:JETP:1969,Chester:PRA:1970,Leggett:PRL:1970}.
Dipolar Bose-Einstein condensates have emerged as a unique platform for the experimental exploration of such superfluid solids \cite{Ferlaino:PRX:2019,Pfau:PRX:2019,Modugno:PRL:2019,Guo:Nature:2019,Tanzi:Nature:2019,Natale:PRL:2019,Tanzi:Science:2021,Petter:PRA:2021,Biagioni:PRX:2022,Bland:PRL:2022,Norcia:PRL:2022}, and have attracted substantial theoretical interest~\cite{Mazzanti:PRL:2017,Guo:Nature:2019,zhang19,Blakie:CTP:2020,ferlaino20,zhang21,Pfau:PRR:2021,Pfau:PRL:2021,wachtler16,Blakie:PRA:2016,Blakie:PRA2:2016,bottcher19} in recent years.

The physics of dipolar supersolids is closely connected to quantum fluctuations~\cite{Pelster:PRA:2011,pelster12} which stabilise the condensate~\cite{Pfau:nature:2016,Pfau:nature2:2016,wachtler16,Blakie:PRA:2016,Blakie:PRA2:2016,Saito:JPhysJ:2016} against dipolar collapse \cite{Lahaye:PRL:2008,Pfau:NatPhys:2008}  due to the attractive part of the meanfield interaction between the dipoles of the atoms.
The important role of quantum fluctuations results from the anisotropic nature of the long-range dipole-dipole interaction, which also gives rise to a range of pattern formation phenomena~\cite{macia12,macia14,zhang19,zhang21,Pfau:PRR:2021,gallemi22,guijarro22,mazzanti23}.

\begin{figure}[t]
\centering
\includegraphics[width=0.6\linewidth]{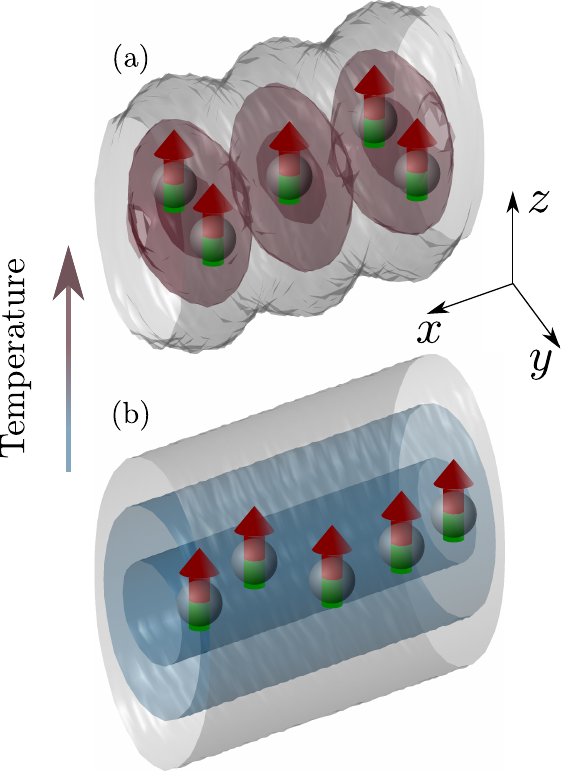}
\caption{
Schematic representation of the dipolar system in the tubular geometry. Dipoles are polarized along the $z$ axis, while the system extends infinitely along the $x$ axis. By keeping the number of condensed atoms fixed and increasing the temperature, the system transitions from an unmodulated gas (b) to a supersolid (a). The trapping strengths in the $y-z$ plane are given by $\omega_{\perp}=0.165 \epsilon_{d}/\hbar$, with $\epsilon_{d} = \hbar^2/(m (12 \pi a_{\rm d})^2)$ and $a_{\rm d}$ denoting the dipolar length.
}
\label{fig_1}
\end{figure}

For the same reasons, thermal fluctuations can also have substantial effects on the phases of dipolar quantum gases, even well below the condensation temperature~\cite{oktel19,oktel20,Ferlaino:PRL:2021,baena22}. In particular, recent calculations showed how heating a dipolar superfluid can induce a transition to a solid phase with a periodically modulated condensate density~\cite{baena22}.

In this work, we use Bogoliubov theory~\cite{oktel19,oktel20} to numerically study finite-temperature effects on a dipolar BEC with strong transverse confinement in the thermodynamic limit (see Fig.~\ref{fig_1}). Recent studies of the zero-temperature phase diagram showed that the superfluid-supersolid phase-transition of this system  can of first- as well as of second-order~\cite{Blakie:CTP:2020,ferlaino20,smith22,Buchler:PRA:2022}. Here, we explore the effects of thermal fluctuations on the nature of the phase transition.

\section{\label{sec:finite_T}Finite temperature theory}
Detailed discussions of finite-temperature effects in dilute Bose Einstein condensates can be found, e.g., in~\cite{Giorgini1997,Griffin:PRB:1996,shi98,cormack12,wang21}.
In order to account for the effect of thermal fluctuations in dipolar  BECs, we apply Bogoliubov theory and use local density approximation~\cite{oktel19,baena22}. This yields a  temperature-dependent extended Gross-Pitaevskii equation (TeGPE) given by
\begin{align}
 \mu \psi({\bf r}) =& \bigg(-\frac{\hbar^2\nabla^2}{2m} + U({\bf r}) + \!\!\int \!{\rm d}{\bf r}^\prime V_{\rm dd}({\bf r}-{\bf r}') \abs{\psi({\bf r}')}^2+ \nonumber\\
 &+ \frac{4\pi\hbar^2 a}{m} \abs{\psi({\bf r})}^2 + H_{\rm qu}({\bf r}) + H_{\rm th}({\bf r}) \bigg) \psi({\bf r}) \label{TeGPE}
\end{align}
for the condensate wave function $\psi({\bf r})$.
Here, $\mu$ is the chemical potential, $m$ is the mass, $U$ describes de trapping potential, $V_{\rm dd}$ denotes the dipolar interactions and $a$ corresponds to the s-wave scattering length. Furthermore, $H_{\rm qu}$ and $H_{\rm th}$ describe effective nonlinear potentials that arise from quantum fluctuations and thermal fluctuations, respectively. They are given by
\begin{align}
     H_{\rm qu}({\bf r}) &= \frac{32}{3 \sqrt{\pi}} g \sqrt{a^3} Q_5(a_{\rm d}/a) \abs{ \psi({\bf r}) }^3 \label{QF} \\
 H_{\rm th}({\bf r}) &= { \int \frac{d{\bf k}}{(2 \pi)^3} \frac{ 1 }{ \left( e^{\beta \varepsilon_{{\bf k}}}-1 \right) } \tilde{V}({\bf k}) \frac{ \tau_{ {\bf k} } }{ \varepsilon_{\bf k}({\bf r}) } } \label{TF} \ ,
\end{align}
where $\varepsilon_{{\bf k}}({\bf r}) = \sqrt{ \tau_{ {\bf k} } \left( \tau_{ {\bf k} } + 2 |\psi({\bf r})|^2 \tilde{V}({\bf k}) \right) }$ is the Bogoliubov excitation spectrum for a given local density $|\psi({\bf r})|^2$ of the BEC, $\tau_{ {\bf k} } = \frac{\hbar^2 k^2}{2m}$, $\beta = 1/k_B T$ and $T$ denotes temperature. $\tilde{V}({\bf k})$ represents the Fourier transform of the sum of the dipole-dipole interaction (DDI) and the  contact interaction, given by
\begin{equation}
\tilde{V}( {\bf k} ) = \frac{4 \pi \hbar^2 a}{m} + \frac{4 \pi \hbar^2 a_{\text{d}}}{m} \left( 3 \frac{k_z^2}{k^2} -1\right) \ .
\label{fourier_int}
\end{equation}
The parameter $a_{\rm d} = m C_{\text{dd}}/(12 \pi \hbar^2)$ corresponds to the dipolar length, $C_{\text{dd}}$ is the strength of the dipolar interaction, and the auxiliary function $Q_5(a_{\rm d}/a)$ is given by~\cite{pelster12}
\begin{align}
 Q_5(a_{\rm d}/a) = \int_{0}^1 du \left( 1 - \frac{a_{\rm d}}{a} + 3 \left( \frac{a_{\rm d}}{a} \right) u^2 \right)^{5/2} \ .
\end{align}

\begin{figure}[t]
\centering
\includegraphics[width=0.9\linewidth]{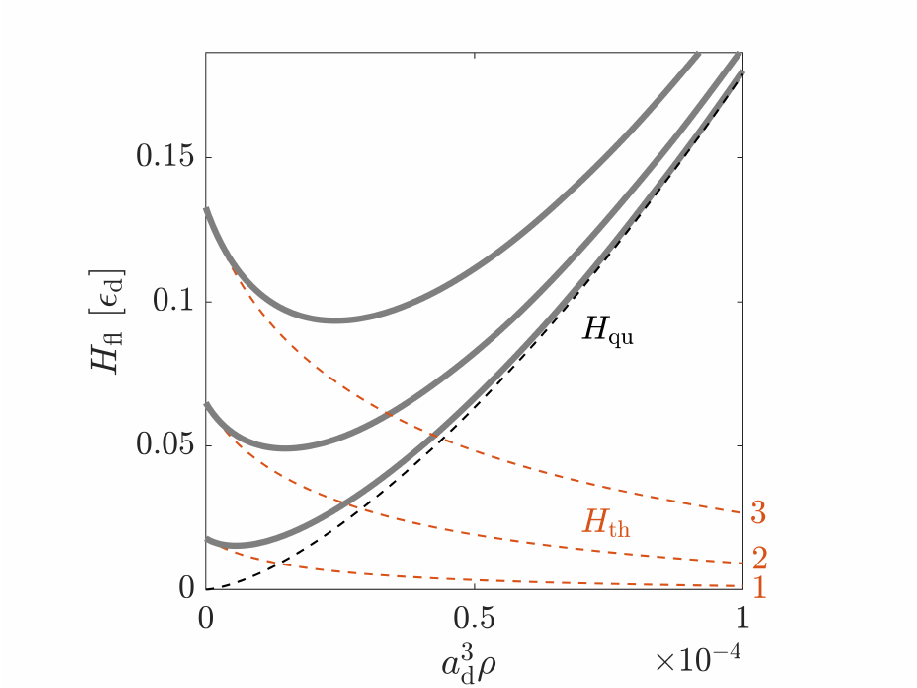}
\caption{Density dependence of the energy contributions to the TeGPE from quantum fluctuations, $H_{\rm qu}$, and thermal fluctuations, $H_{\rm th}$, along with the total energy $H_{\rm fl}=H_{\rm th}+H_{\rm qu}$. The results are shown for $a/a_{\rm d} = 0.7$ and different indicated temperatures $k_B T/\epsilon_d = 1,2,3$.}
\label{fig_2}
\end{figure}

The term in Eq.~\ref{QF} that accounts for quantum fluctuations acts as a defocusing nonlinearity whose strength increases with the condensed density and is responsible for arresting collapse as discussed above ~\cite{kadau16,ferlaino16,wachtler16}. Thermal fluctuations, as described by Eq.~\ref{TF}, on the other hand, generate a focusing
nonlinearity that decreases with increasing density $\rho$~\cite{oktel19,baena22}.
Fig.~\ref{fig_2} illustrates this density dependence and shows that the total fluctuation energy, $H_{\rm fl}=H_{\rm qu}+H_{\rm th}$, features a minimum that shifts towards higher densities with increasing temperature.

The evaluation of Eq.~\ref{TF} for values of the scattering length that are lower than the dipole length requires special attention since the integrand can in this case become complex. This reflects the instability of a homogeneous condensate, as the excitation spectrum, $\varepsilon_{{\bf k}}$, turns imaginary for small momenta and $a < a_{\rm d}$.
The finite transverse size of the partially confined condensate, however, introduces a natural low-momentum cut-off for the considered system. Due to the symmetry of the dipole-dipole interaction, the contribution to the fluctuation energies depends only on $k_z$ and $k_{\rho} = \sqrt{k_x^2 + k_y^2}$. Considering radial confinement as shown in Fig.\ref{fig_1} with typical system sizes $l_y$ and $l_z$ along the $y$-axis and the $z$-axis, respectively, one obtains lower bounds, $k_z>2\pi/l_z$ and $k_\rho>2\pi/l_y$, for both  momenta. Here, we use $k_z> 0.007/a_{\rm d}$ and $k_{\rho}>0.017/ a_{\rm d}$ and have checked that a $30\%$ increase of these values does not significantly affect the numerical results.
\section{\label{sec:results}Finite-temperature Phase-diagram}
\begin{figure}[b]
\centering
\includegraphics[width=1\linewidth]{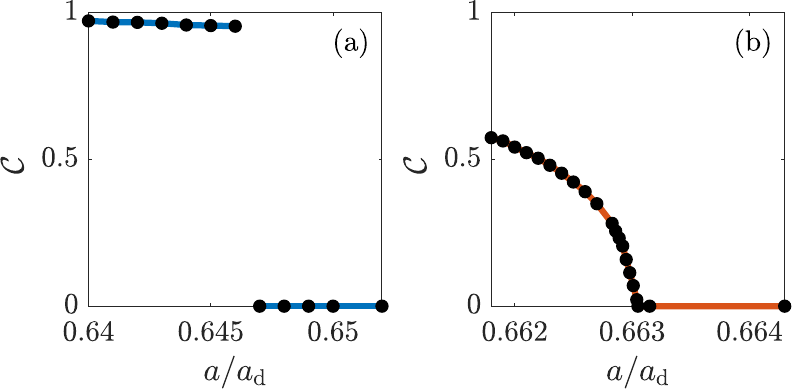}
\caption{{Contrast of the wave function versus scattering length in a (a) first and (b) second order phase transition {for $\bar{\rho} a_{\rm d}=4.77$ and two different temperatures: $k_B T/\epsilon_d=0$ (a) and $k_B T/\epsilon_d=2$ (b)}}.}
\label{fig_3}
\end{figure}

Equation~\ref{TeGPE} can be solved numerically by imaginary time evolution to obtain the condensate wave function $\psi$ for a finite temperature, $T$, and a fixed condensate density or chemical potential, $\mu$. Figure~\ref{fig_3} shows the contrast
\begin{equation}
\mathcal{C} = \frac{\rho_{\text{max}}-\rho_{\text{min}}}{\rho_{\text{max}}+\rho_{\text{min}} }
\end{equation}
where $\rho_{\rm max}$ and $\rho_{\rm min}$ denote the maximum and minimum of the axial density $\rho(x) = \int dy dz \abs{ \Psi({\bf r}) }^2$ along the $x$-axis. The latter is also used to define the overall axial density $\bar{\rho}=L^{-1}\int_0^L{\rm d}x \rho(x)$ for a given value of the length $L$ of the simulation box along the $x$-direction.

The axial density contrast vanishes in the superfluid phase, for large ratios $a/a_{\rm d}$ in Fig.\ref{fig_3}, and takes on finite values below a critical value of $a/a_{\rm d}$ as one enters the solid phase with finite density modulations. For the chosen density of $\bar{\rho} a_{\rm d}=4.77$, the zero temperature simulation yields a discontinuous increase of the contrast, characteristic for a first order phase transition. For a finite temperature of $k_B T/\epsilon_d=2$, however, one finds a second order phase transition with a continuous rise of the density contrast. Apart from shifting the transition point, thermal fluctuations, therefore, may also qualitatively affect the fluid-solid phase transition.

\begin{figure}[t]
\centering
\includegraphics[width=1\linewidth]{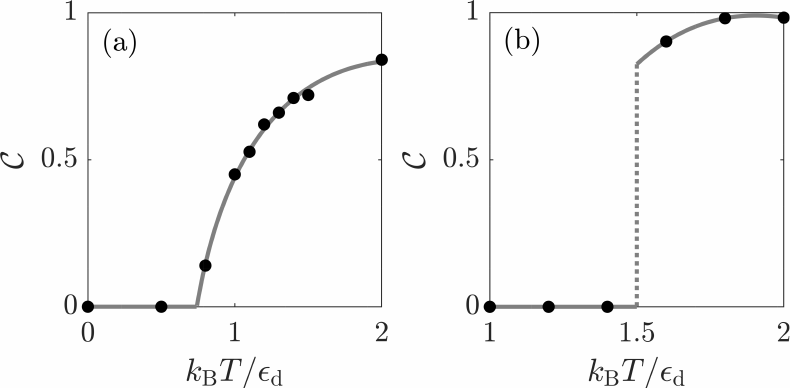}
\caption{{Contrast of the wave function as a function of temperature for $\bar{\rho} a_{\rm d}=6.89$, $a/a_{\rm d} = 0.676$ (a) and {$\bar{\rho} a_{\rm d}=4$, $a/a_{\rm d} = 0.64$ (b)}. The lines correspond to eye-guides.}}
\label{fig_4}
\end{figure}

Thermal effects are further illustrated in Fig.~\ref{fig_4}, where we show the contrast across the phase transition as a function of temperature for two different values of the condensate density. In all depicted cases, heating the system induces a transition to a density-modulated state, irrespective of the order of the transition. Around the second order phase transition, one finds moderately modulated states with a density contrast that remains significantly below unity. Concurrently, such states are expected to feature a substantial superfluid fraction ~\cite{legget70,sepulveda08,leggett98} and should, therefore, realise a supersolid. On the other hand, a direct first order transition from a superfluid to a solid with near unit modulation contrast and no global superfluidity should eventually occur with decreasing density.

\begin{figure}[t]
\centering
\includegraphics[width=1\linewidth]{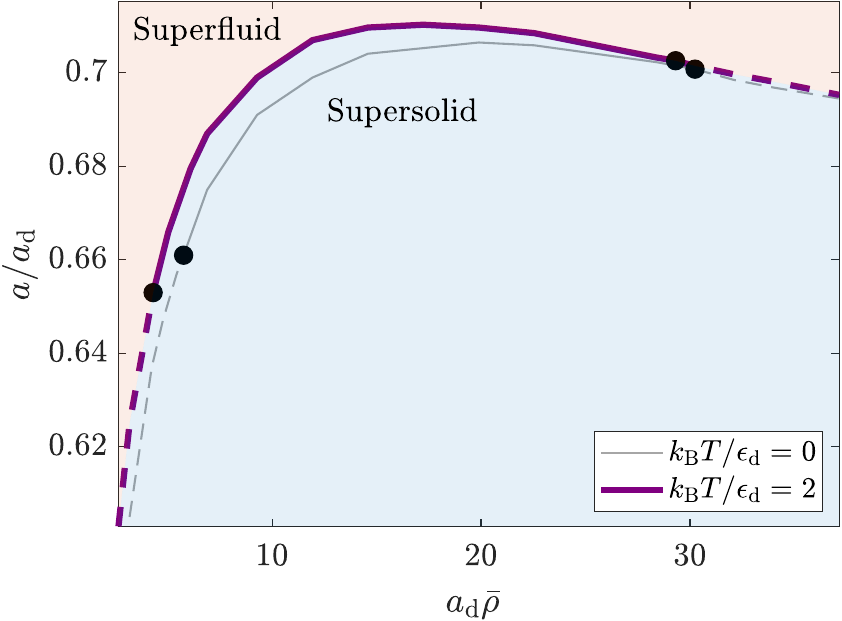}
\caption{Superfluid-supersolid phase-diagram for $T=0$ (black line) and $k_\mathrm{B} T/\epsilon_\mathrm{d}=2$ (purple line). Solid lines show regions where the transition is of continuous or second order whereas dashed lines indicate the presence of a first order phase transition. The black point marks the low and high density critical point separating regions of first- and second-order phase-transitions.
}
\label{fig_5}
\end{figure}

Figure~\ref{fig_5} provides a more complete picture of the fluid-solid transition, showing the phase diagram at zero temperature and $k_B T/\epsilon_d = 2$ as a function of the density and the competing interaction strengths. The chosen parameters lie in typical regimes of current experiments, e.g., whereby the temperature corresponds to $T \simeq 87$nK for a quantum gas of $^{164}$Dy atoms. The calculations show that such low temperatures do not qualitatively alter the phase diagram compared to the ground state behaviour of dipolar condensates, discussed recently in \cite{ferlaino20,smith22}.

As the temperature is increased, the solid-fluid transition line shifts towards larger values of the scattering length $a$. Starting from the superfluid phase close to the quantum phase transition ($T=0$) and increasing the temperature, therefore, leads to the emergence of a solid phase upon heating the system regardless of the precise values of the otherwise fixed parameters (i.e., $a$, $a_{\rm d}$, and $\bar{\rho}$). This effect, which has been reported in recent experiments with $^{164}$Dy atoms~\cite{ferlaino21, baena22}, can be understood from the characteristic density dependence of the energy, $H_{\rm th}$, of thermal fluctuations shown in Fig.~\ref{fig_2}. A decreasing energy with increasing density, $|\psi|^2$, implies that $H_{\rm th}[|\psi({\bf r})|]$ acts as a focusing nonlinearity in the generalized GPE \cite{baena22} and, therefore, tends to support the  density-modulated phase.

\begin{figure}[t]
\centering
\includegraphics[width=1\linewidth]{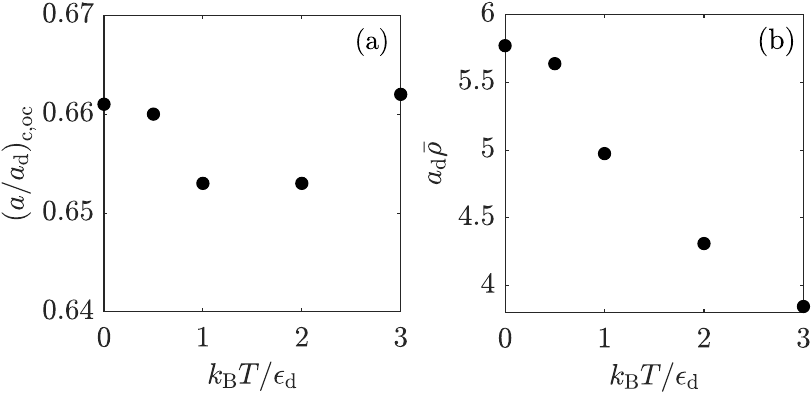}
\caption{{Scattering length (a) and density (b) of the low density critical point as a function of the temperature.
}}
\label{fig_7}
\end{figure}

One observes in Fig.~\ref{fig_5} a convergence of the phase boundaries for the two different temperatures for large densities, which shows again that thermal effects on the phase boundary weaken as the density increases. This can again be readily understood from Fig.~\ref{fig_2}, which shows that quantum fluctuations yield the dominant contribution to the energy correction $H_{\rm fl}$ at higher densities. While thermal fluctuations always shift the phase boundary towards larger scattering lengths, $a$, their effect on the critical density depends on the density itself. At lower densities, where the energy corrections from quantum fluctuations and thermal fluctuations are comparable, the phase boundary is shifted towards lower densities and thereby facilitates the formation of the solid phase. On the contrary, at higher densities, where quantum fluctuations dominate the energy correction, $H_{\rm fl}$, a larger temperature requires an increased density to form a modulated state. Yet, heating still facilitates the solid phase, since the critical scattering length decreases with density in this regime (see Fig.\ref{fig_2}). This effect is illustrated in Fig.\ref{fig_7}, where we show the contrast as a function of $\bar{\rho}$ in the two different density regimes.

We finally discuss the order of the phase transition and how it is affected by thermal fluctuations. At very low densities, the transition is of first-order type but turns into a continuous second order phase transition with increasing density. Eventually, the phase transition becomes once again discontinuous in the high density regime. This general phenomenology of the quantum phase transition ($T=0$) \cite{ferlaino20,smith22} prevails at finite temperatures, while thermal fluctuations can shift the critical points at which the order of the phase transition changes.

\begin{figure}[t]
\centering
\includegraphics[width=1\linewidth]{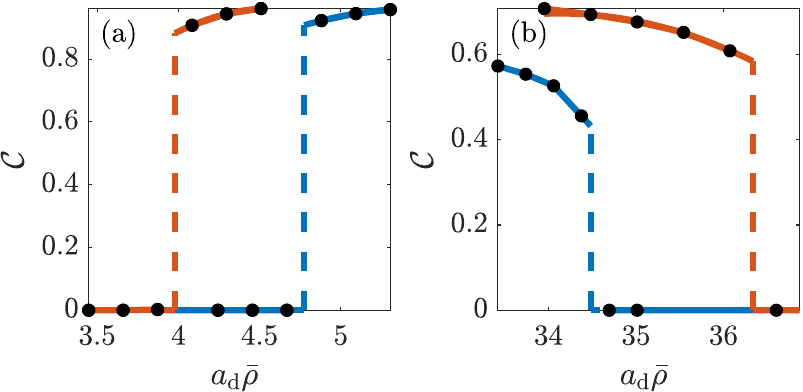}
\caption{
Contrast $\mathcal C$ of the wave function versus axial density in the (a) low density and (b) high density regimes as function of the density. }
\label{fig_6}
\end{figure}

At the low density critical point (cf. Fig~\ref{fig_5}) the change of the critical value of the scattering length ($(a/a_{\rm dd})=0.655\pm 0.005$) is small compared to the experimental resolution for the values considered [cf. Fig.~\ref{fig_7}(a)]. In contrast to that, one finds a substantial effect on the critical density, which decreases significantly with increasing temperature of the BEC [cf. Fig.~\ref{fig_7}(b)]. On the other hand, the shift in the critical point is less pronounced in the high density regime, as can be seen in Fig.~\ref{fig_5}. This is a consequence of the aforementioned weakening of thermal effects for increasing condensate density.

\section{\label{sec:conclusions}Conclusions}

In conclusion, we have characterized finite-temperature effects in the phase diagram of an elongated dipolar BEC. We have shown that an increase of temperature at constant condensed density yields a transition from an unmodulated superfluid to a supersolid, thus pushing the solid-fluid boundaries of the zero temperature diagram to higher values of the scattering length. For a sufficiently large condensed density we enter
a regime where quantum fluctuations dominate and thermal effects become small in comparison. We have also seen how the low density critical point where the fluid-solid transition shifts from first to second order (or vice versa) changes with temperature and have shown that it moves to lower values of the condensed density, thus yielding a range of parameters for which temperature effectively changes the order of the phase transition. We have observed that the high density critical point experiments a less significant shift than the low density one, due to the weakening of thermal effects as the axial condensed density increases.

The role of temperature in dipolar systems still remains a relatively unexplored subject. It remains unclear how temperature will affect the different geometrical phases both in infinite and trapped quasi-2D dipolar systems~\cite{zhang19,zhang21,hertkorn21} as well as their superfluid properties~\cite{norcia21}. {Furthermore}, improved theoretical calculations for specific geometries where the local density approximation {is not necessary could present avenues towards} more accurate quantitative predictions 
{that can be}
compared with future experiments. Similarly, the realization of ab-initio calculations~\cite{bottcher19} that are able to fully account for the effect of temperature could help extending the formalism presented in this work to the high temperature regime, where the fraction of condensed atoms is small.

\acknowledgements

This work was supported by the DNRF through the Center of Excellence
”CCQ” (Grant agreement no.: DNRF156) and the Carlsberg Foundation through the ’Semper Ardens’ Research
Project QCooL. F.M. acknowledges funding from the Ministerio de Economía y Competitividad (PID2021-128910NB-100 AEI/FEDER UE).

\bibliography{dipole_tube_T}

\begin{thebibliography}{56}%
\makeatletter
\providecommand \@ifxundefined [1]{%
 \@ifx{#1\undefined}
}%
\providecommand \@ifnum [1]{%
 \ifnum #1\expandafter \@firstoftwo
 \else \expandafter \@secondoftwo
 \fi
}%
\providecommand \@ifx [1]{%
 \ifx #1\expandafter \@firstoftwo
 \else \expandafter \@secondoftwo
 \fi
}%
\providecommand \natexlab [1]{#1}%
\providecommand \enquote  [1]{``#1''}%
\providecommand \bibnamefont  [1]{#1}%
\providecommand \bibfnamefont [1]{#1}%
\providecommand \citenamefont [1]{#1}%
\providecommand \href@noop [0]{\@secondoftwo}%
\providecommand \href [0]{\begingroup \@sanitize@url \@href}%
\providecommand \@href[1]{\@@startlink{#1}\@@href}%
\providecommand \@@href[1]{\endgroup#1\@@endlink}%
\providecommand \@sanitize@url [0]{\catcode `\\12\catcode `\$12\catcode
  `\&12\catcode `\#12\catcode `\^12\catcode `\_12\catcode `\%12\relax}%
\providecommand \@@startlink[1]{}%
\providecommand \@@endlink[0]{}%
\providecommand \url  [0]{\begingroup\@sanitize@url \@url }%
\providecommand \@url [1]{\endgroup\@href {#1}{\urlprefix }}%
\providecommand \urlprefix  [0]{URL }%
\providecommand \Eprint [0]{\href }%
\providecommand \doibase [0]{http://dx.doi.org/}%
\providecommand \selectlanguage [0]{\@gobble}%
\providecommand \bibinfo  [0]{\@secondoftwo}%
\providecommand \bibfield  [0]{\@secondoftwo}%
\providecommand \translation [1]{[#1]}%
\providecommand \BibitemOpen [0]{}%
\providecommand \bibitemStop [0]{}%
\providecommand \bibitemNoStop [0]{.\EOS\space}%
\providecommand \EOS [0]{\spacefactor3000\relax}%
\providecommand \BibitemShut  [1]{\csname bibitem#1\endcsname}%
\let\auto@bib@innerbib\@empty
\bibitem [{\citenamefont {Andreev}\ and\ \citenamefont
  {Lifshitz}(1969)}]{Andreev:JETP:1969}%
  \BibitemOpen
  \bibfield  {author} {\bibinfo {author} {\bibfnamefont {A.~F}\ \bibnamefont
  {Andreev}}\ and\ \bibinfo {author} {\bibfnamefont {I.~M.}\ \bibnamefont
  {Lifshitz}},\ }\bibfield  {title} {\enquote {\bibinfo {title} {Quantum theory
  of defects in crystals},}\ }\href@noop {} {\bibfield  {journal} {\bibinfo
  {journal} {Sov. Phys. JETP}\ }\textbf {\bibinfo {volume} {29}},\ \bibinfo
  {pages} {1107--1113} (\bibinfo {year} {1969})}\BibitemShut {NoStop}%
\bibitem [{\citenamefont {Chester}(1970)}]{Chester:PRA:1970}%
  \BibitemOpen
  \bibfield  {author} {\bibinfo {author} {\bibfnamefont {G.~V.}\ \bibnamefont
  {Chester}},\ }\bibfield  {title} {\enquote {\bibinfo {title} {Speculations on
  bose-einstein condensation and quantum crystals},}\ }\href@noop {} {\bibfield
   {journal} {\bibinfo  {journal} {Phys. Rev. A}\ }\textbf {\bibinfo {volume}
  {2}},\ \bibinfo {pages} {256--258} (\bibinfo {year} {1970})}\BibitemShut
  {NoStop}%
\bibitem [{\citenamefont {Leggett}(1970{\natexlab{a}})}]{Leggett:PRL:1970}%
  \BibitemOpen
  \bibfield  {author} {\bibinfo {author} {\bibfnamefont {A.~J.}\ \bibnamefont
  {Leggett}},\ }\bibfield  {title} {\enquote {\bibinfo {title} {Can a solid be
  “superfluid”?}}\ }\href@noop {} {\bibfield  {journal} {\bibinfo
  {journal} {Phys. Rev. Lett.}\ }\textbf {\bibinfo {volume} {25}},\ \bibinfo
  {pages} {1543--1546} (\bibinfo {year} {1970}{\natexlab{a}})}\BibitemShut
  {NoStop}%
\bibitem [{\citenamefont {Chomaz}\ \emph {et~al.}(2019)\citenamefont {Chomaz},
  \citenamefont {Petter}, \citenamefont {Ilzh\"ofer}, \citenamefont {Natale},
  \citenamefont {Trautmann}, \citenamefont {Politi}, \citenamefont
  {Durastante}, \citenamefont {van Bijnen}, \citenamefont {Patscheider},
  \citenamefont {Sohmen}, \citenamefont {Mark},\ and\ \citenamefont
  {Ferlaino}}]{Ferlaino:PRX:2019}%
  \BibitemOpen
  \bibfield  {author} {\bibinfo {author} {\bibfnamefont {L.}~\bibnamefont
  {Chomaz}}, \bibinfo {author} {\bibfnamefont {D.}~\bibnamefont {Petter}},
  \bibinfo {author} {\bibfnamefont {P.}~\bibnamefont {Ilzh\"ofer}}, \bibinfo
  {author} {\bibfnamefont {G.}~\bibnamefont {Natale}}, \bibinfo {author}
  {\bibfnamefont {A.}~\bibnamefont {Trautmann}}, \bibinfo {author}
  {\bibfnamefont {C.}~\bibnamefont {Politi}}, \bibinfo {author} {\bibfnamefont
  {G.}~\bibnamefont {Durastante}}, \bibinfo {author} {\bibfnamefont {R.~M.~W.}\
  \bibnamefont {van Bijnen}}, \bibinfo {author} {\bibfnamefont
  {A.}~\bibnamefont {Patscheider}}, \bibinfo {author} {\bibfnamefont
  {M.}~\bibnamefont {Sohmen}}, \bibinfo {author} {\bibfnamefont {M.~J.}\
  \bibnamefont {Mark}}, \ and\ \bibinfo {author} {\bibfnamefont
  {F.}~\bibnamefont {Ferlaino}},\ }\bibfield  {title} {\enquote {\bibinfo
  {title} {Long-lived and transient supersolid behaviors in dipolar quantum
  gases},}\ }\href {\doibase 10.1103/PhysRevX.9.021012} {\bibfield  {journal}
  {\bibinfo  {journal} {Phys. Rev. X}\ }\textbf {\bibinfo {volume} {9}},\
  \bibinfo {pages} {021012} (\bibinfo {year} {2019})}\BibitemShut {NoStop}%
\bibitem [{\citenamefont {B\"ottcher}\ \emph
  {et~al.}(2019{\natexlab{a}})\citenamefont {B\"ottcher}, \citenamefont
  {Schmidt}, \citenamefont {Wenzel}, \citenamefont {Hertkorn}, \citenamefont
  {Guo}, \citenamefont {Langen},\ and\ \citenamefont {Pfau}}]{Pfau:PRX:2019}%
  \BibitemOpen
  \bibfield  {author} {\bibinfo {author} {\bibfnamefont {Fabian}\ \bibnamefont
  {B\"ottcher}}, \bibinfo {author} {\bibfnamefont {Jan-Niklas}\ \bibnamefont
  {Schmidt}}, \bibinfo {author} {\bibfnamefont {Matthias}\ \bibnamefont
  {Wenzel}}, \bibinfo {author} {\bibfnamefont {Jens}\ \bibnamefont {Hertkorn}},
  \bibinfo {author} {\bibfnamefont {Mingyang}\ \bibnamefont {Guo}}, \bibinfo
  {author} {\bibfnamefont {Tim}\ \bibnamefont {Langen}}, \ and\ \bibinfo
  {author} {\bibfnamefont {Tilman}\ \bibnamefont {Pfau}},\ }\bibfield  {title}
  {\enquote {\bibinfo {title} {Transient supersolid properties in an array of
  dipolar quantum droplets},}\ }\href {\doibase 10.1103/PhysRevX.9.011051}
  {\bibfield  {journal} {\bibinfo  {journal} {Phys. Rev. X}\ }\textbf {\bibinfo
  {volume} {9}},\ \bibinfo {pages} {011051} (\bibinfo {year}
  {2019}{\natexlab{a}})}\BibitemShut {NoStop}%
\bibitem [{\citenamefont {Tanzi}\ \emph
  {et~al.}(2019{\natexlab{a}})\citenamefont {Tanzi}, \citenamefont {Lucioni},
  \citenamefont {Fam\`a}, \citenamefont {Catani}, \citenamefont {Fioretti},
  \citenamefont {Gabbanini}, \citenamefont {Bisset}, \citenamefont {Santos},\
  and\ \citenamefont {Modugno}}]{Modugno:PRL:2019}%
  \BibitemOpen
  \bibfield  {author} {\bibinfo {author} {\bibfnamefont {L.}~\bibnamefont
  {Tanzi}}, \bibinfo {author} {\bibfnamefont {E.}~\bibnamefont {Lucioni}},
  \bibinfo {author} {\bibfnamefont {F.}~\bibnamefont {Fam\`a}}, \bibinfo
  {author} {\bibfnamefont {J.}~\bibnamefont {Catani}}, \bibinfo {author}
  {\bibfnamefont {A.}~\bibnamefont {Fioretti}}, \bibinfo {author}
  {\bibfnamefont {C.}~\bibnamefont {Gabbanini}}, \bibinfo {author}
  {\bibfnamefont {R.~N.}\ \bibnamefont {Bisset}}, \bibinfo {author}
  {\bibfnamefont {L.}~\bibnamefont {Santos}}, \ and\ \bibinfo {author}
  {\bibfnamefont {G.}~\bibnamefont {Modugno}},\ }\bibfield  {title} {\enquote
  {\bibinfo {title} {Observation of a dipolar quantum gas with metastable
  supersolid properties},}\ }\href {\doibase 10.1103/PhysRevLett.122.130405}
  {\bibfield  {journal} {\bibinfo  {journal} {Phys. Rev. Lett.}\ }\textbf
  {\bibinfo {volume} {122}},\ \bibinfo {pages} {130405} (\bibinfo {year}
  {2019}{\natexlab{a}})}\BibitemShut {NoStop}%
\bibitem [{\citenamefont {Guo}\ \emph {et~al.}(2019)\citenamefont {Guo},
  \citenamefont {B{\"o}ttcher}, \citenamefont {Hertkorn}, \citenamefont
  {Schmidt}, \citenamefont {Wenzel}, \citenamefont {B{\"u}chler}, \citenamefont
  {Langen},\ and\ \citenamefont {Pfau}}]{Guo:Nature:2019}%
  \BibitemOpen
  \bibfield  {author} {\bibinfo {author} {\bibfnamefont {Mingyang}\
  \bibnamefont {Guo}}, \bibinfo {author} {\bibfnamefont {Fabian}\ \bibnamefont
  {B{\"o}ttcher}}, \bibinfo {author} {\bibfnamefont {Jens}\ \bibnamefont
  {Hertkorn}}, \bibinfo {author} {\bibfnamefont {Jan-Niklas}\ \bibnamefont
  {Schmidt}}, \bibinfo {author} {\bibfnamefont {Matthias}\ \bibnamefont
  {Wenzel}}, \bibinfo {author} {\bibfnamefont {Hans~Peter}\ \bibnamefont
  {B{\"u}chler}}, \bibinfo {author} {\bibfnamefont {Tim}\ \bibnamefont
  {Langen}}, \ and\ \bibinfo {author} {\bibfnamefont {Tilman}\ \bibnamefont
  {Pfau}},\ }\bibfield  {title} {\enquote {\bibinfo {title} {The low-energy
  goldstone mode in a trapped dipolar supersolid},}\ }\href {\doibase
  10.1038/s41586-019-1569-5} {\bibfield  {journal} {\bibinfo  {journal}
  {Nature}\ }\textbf {\bibinfo {volume} {574}},\ \bibinfo {pages} {386--389}
  (\bibinfo {year} {2019})}\BibitemShut {NoStop}%
\bibitem [{\citenamefont {Tanzi}\ \emph
  {et~al.}(2019{\natexlab{b}})\citenamefont {Tanzi}, \citenamefont {Roccuzzo},
  \citenamefont {Lucioni}, \citenamefont {Fam{\`a}}, \citenamefont {Fioretti},
  \citenamefont {Gabbanini}, \citenamefont {Modugno}, \citenamefont {Recati},\
  and\ \citenamefont {Stringari}}]{Tanzi:Nature:2019}%
  \BibitemOpen
  \bibfield  {author} {\bibinfo {author} {\bibfnamefont {L.}~\bibnamefont
  {Tanzi}}, \bibinfo {author} {\bibfnamefont {S.~M.}\ \bibnamefont {Roccuzzo}},
  \bibinfo {author} {\bibfnamefont {E.}~\bibnamefont {Lucioni}}, \bibinfo
  {author} {\bibfnamefont {F.}~\bibnamefont {Fam{\`a}}}, \bibinfo {author}
  {\bibfnamefont {A.}~\bibnamefont {Fioretti}}, \bibinfo {author}
  {\bibfnamefont {C.}~\bibnamefont {Gabbanini}}, \bibinfo {author}
  {\bibfnamefont {G.}~\bibnamefont {Modugno}}, \bibinfo {author} {\bibfnamefont
  {A.}~\bibnamefont {Recati}}, \ and\ \bibinfo {author} {\bibfnamefont
  {S.}~\bibnamefont {Stringari}},\ }\bibfield  {title} {\enquote {\bibinfo
  {title} {Supersolid symmetry breaking from compressional oscillations in a
  dipolar quantum gas},}\ }\href {\doibase 10.1038/s41586-019-1568-6}
  {\bibfield  {journal} {\bibinfo  {journal} {Nature}\ }\textbf {\bibinfo
  {volume} {574}},\ \bibinfo {pages} {382--385} (\bibinfo {year}
  {2019}{\natexlab{b}})}\BibitemShut {NoStop}%
\bibitem [{\citenamefont {Natale}\ \emph {et~al.}(2019)\citenamefont {Natale},
  \citenamefont {van Bijnen}, \citenamefont {Patscheider}, \citenamefont
  {Petter}, \citenamefont {Mark}, \citenamefont {Chomaz},\ and\ \citenamefont
  {Ferlaino}}]{Natale:PRL:2019}%
  \BibitemOpen
  \bibfield  {author} {\bibinfo {author} {\bibfnamefont {G.}~\bibnamefont
  {Natale}}, \bibinfo {author} {\bibfnamefont {R.~M.~W.}\ \bibnamefont {van
  Bijnen}}, \bibinfo {author} {\bibfnamefont {A.}~\bibnamefont {Patscheider}},
  \bibinfo {author} {\bibfnamefont {D.}~\bibnamefont {Petter}}, \bibinfo
  {author} {\bibfnamefont {M.~J.}\ \bibnamefont {Mark}}, \bibinfo {author}
  {\bibfnamefont {L.}~\bibnamefont {Chomaz}}, \ and\ \bibinfo {author}
  {\bibfnamefont {F.}~\bibnamefont {Ferlaino}},\ }\bibfield  {title} {\enquote
  {\bibinfo {title} {Excitation spectrum of a trapped dipolar supersolid and
  its experimental evidence},}\ }\href {\doibase
  10.1103/PhysRevLett.123.050402} {\bibfield  {journal} {\bibinfo  {journal}
  {Phys. Rev. Lett.}\ }\textbf {\bibinfo {volume} {123}},\ \bibinfo {pages}
  {050402} (\bibinfo {year} {2019})}\BibitemShut {NoStop}%
\bibitem [{\citenamefont {Tanzi}\ \emph {et~al.}(2021)\citenamefont {Tanzi},
  \citenamefont {Maloberti}, \citenamefont {Biagioni}, \citenamefont
  {Fioretti}, \citenamefont {Gabbanini},\ and\ \citenamefont
  {Modugno}}]{Tanzi:Science:2021}%
  \BibitemOpen
  \bibfield  {author} {\bibinfo {author} {\bibfnamefont {L.}~\bibnamefont
  {Tanzi}}, \bibinfo {author} {\bibfnamefont {J.~G.}\ \bibnamefont
  {Maloberti}}, \bibinfo {author} {\bibfnamefont {G.}~\bibnamefont {Biagioni}},
  \bibinfo {author} {\bibfnamefont {A.}~\bibnamefont {Fioretti}}, \bibinfo
  {author} {\bibfnamefont {C.}~\bibnamefont {Gabbanini}}, \ and\ \bibinfo
  {author} {\bibfnamefont {G.}~\bibnamefont {Modugno}},\ }\bibfield  {title}
  {\enquote {\bibinfo {title} {Evidence of superfluidity in a dipolar
  supersolid from nonclassical rotational inertia},}\ }\href {\doibase
  10.1126/science.aba4309} {\bibfield  {journal} {\bibinfo  {journal}
  {Science}\ }\textbf {\bibinfo {volume} {371}},\ \bibinfo {pages} {1162--1165}
  (\bibinfo {year} {2021})}\BibitemShut {NoStop}%
\bibitem [{\citenamefont {Petter}\ \emph {et~al.}(2021)\citenamefont {Petter},
  \citenamefont {Patscheider}, \citenamefont {Natale}, \citenamefont {Mark},
  \citenamefont {Baranov}, \citenamefont {van Bijnen}, \citenamefont
  {Roccuzzo}, \citenamefont {Recati}, \citenamefont {Blakie}, \citenamefont
  {Baillie}, \citenamefont {Chomaz},\ and\ \citenamefont
  {Ferlaino}}]{Petter:PRA:2021}%
  \BibitemOpen
  \bibfield  {author} {\bibinfo {author} {\bibfnamefont {D.}~\bibnamefont
  {Petter}}, \bibinfo {author} {\bibfnamefont {A.}~\bibnamefont {Patscheider}},
  \bibinfo {author} {\bibfnamefont {G.}~\bibnamefont {Natale}}, \bibinfo
  {author} {\bibfnamefont {M.~J.}\ \bibnamefont {Mark}}, \bibinfo {author}
  {\bibfnamefont {M.~A.}\ \bibnamefont {Baranov}}, \bibinfo {author}
  {\bibfnamefont {R.}~\bibnamefont {van Bijnen}}, \bibinfo {author}
  {\bibfnamefont {S.~M.}\ \bibnamefont {Roccuzzo}}, \bibinfo {author}
  {\bibfnamefont {A.}~\bibnamefont {Recati}}, \bibinfo {author} {\bibfnamefont
  {B.}~\bibnamefont {Blakie}}, \bibinfo {author} {\bibfnamefont
  {D.}~\bibnamefont {Baillie}}, \bibinfo {author} {\bibfnamefont
  {L.}~\bibnamefont {Chomaz}}, \ and\ \bibinfo {author} {\bibfnamefont
  {F.}~\bibnamefont {Ferlaino}},\ }\bibfield  {title} {\enquote {\bibinfo
  {title} {Bragg scattering of an ultracold dipolar gas across the phase
  transition from bose-einstein condensate to supersolid in the free-particle
  regime},}\ }\href {\doibase 10.1103/PhysRevA.104.L011302} {\bibfield
  {journal} {\bibinfo  {journal} {Phys. Rev. A}\ }\textbf {\bibinfo {volume}
  {104}},\ \bibinfo {pages} {L011302} (\bibinfo {year} {2021})}\BibitemShut
  {NoStop}%
\bibitem [{\citenamefont {Biagioni}\ \emph {et~al.}(2022)\citenamefont
  {Biagioni}, \citenamefont {Antolini}, \citenamefont {Ala\~na}, \citenamefont
  {Modugno}, \citenamefont {Fioretti}, \citenamefont {Gabbanini}, \citenamefont
  {Tanzi},\ and\ \citenamefont {Modugno}}]{Biagioni:PRX:2022}%
  \BibitemOpen
  \bibfield  {author} {\bibinfo {author} {\bibfnamefont {Giulio}\ \bibnamefont
  {Biagioni}}, \bibinfo {author} {\bibfnamefont {Nicol\`o}\ \bibnamefont
  {Antolini}}, \bibinfo {author} {\bibfnamefont {Aitor}\ \bibnamefont
  {Ala\~na}}, \bibinfo {author} {\bibfnamefont {Michele}\ \bibnamefont
  {Modugno}}, \bibinfo {author} {\bibfnamefont {Andrea}\ \bibnamefont
  {Fioretti}}, \bibinfo {author} {\bibfnamefont {Carlo}\ \bibnamefont
  {Gabbanini}}, \bibinfo {author} {\bibfnamefont {Luca}\ \bibnamefont {Tanzi}},
  \ and\ \bibinfo {author} {\bibfnamefont {Giovanni}\ \bibnamefont {Modugno}},\
  }\bibfield  {title} {\enquote {\bibinfo {title} {Dimensional crossover in the
  superfluid-supersolid quantum phase transition},}\ }\href {\doibase
  10.1103/PhysRevX.12.021019} {\bibfield  {journal} {\bibinfo  {journal} {Phys.
  Rev. X}\ }\textbf {\bibinfo {volume} {12}},\ \bibinfo {pages} {021019}
  (\bibinfo {year} {2022})}\BibitemShut {NoStop}%
\bibitem [{\citenamefont {Bland}\ \emph {et~al.}(2022)\citenamefont {Bland},
  \citenamefont {Poli}, \citenamefont {Politi}, \citenamefont {Klaus},
  \citenamefont {Norcia}, \citenamefont {Ferlaino}, \citenamefont {Santos},\
  and\ \citenamefont {Bisset}}]{Bland:PRL:2022}%
  \BibitemOpen
  \bibfield  {author} {\bibinfo {author} {\bibfnamefont {T.}~\bibnamefont
  {Bland}}, \bibinfo {author} {\bibfnamefont {E.}~\bibnamefont {Poli}},
  \bibinfo {author} {\bibfnamefont {C.}~\bibnamefont {Politi}}, \bibinfo
  {author} {\bibfnamefont {L.}~\bibnamefont {Klaus}}, \bibinfo {author}
  {\bibfnamefont {M.~A.}\ \bibnamefont {Norcia}}, \bibinfo {author}
  {\bibfnamefont {F.}~\bibnamefont {Ferlaino}}, \bibinfo {author}
  {\bibfnamefont {L.}~\bibnamefont {Santos}}, \ and\ \bibinfo {author}
  {\bibfnamefont {R.~N.}\ \bibnamefont {Bisset}},\ }\bibfield  {title}
  {\enquote {\bibinfo {title} {Two-dimensional supersolid formation in dipolar
  condensates},}\ }\href {\doibase 10.1103/PhysRevLett.128.195302} {\bibfield
  {journal} {\bibinfo  {journal} {Phys. Rev. Lett.}\ }\textbf {\bibinfo
  {volume} {128}},\ \bibinfo {pages} {195302} (\bibinfo {year}
  {2022})}\BibitemShut {NoStop}%
\bibitem [{\citenamefont {Norcia}\ \emph {et~al.}(2022)\citenamefont {Norcia},
  \citenamefont {Poli}, \citenamefont {Politi}, \citenamefont {Klaus},
  \citenamefont {Bland}, \citenamefont {Mark}, \citenamefont {Santos},
  \citenamefont {Bisset},\ and\ \citenamefont {Ferlaino}}]{Norcia:PRL:2022}%
  \BibitemOpen
  \bibfield  {author} {\bibinfo {author} {\bibfnamefont {Matthew~A.}\
  \bibnamefont {Norcia}}, \bibinfo {author} {\bibfnamefont {Elena}\
  \bibnamefont {Poli}}, \bibinfo {author} {\bibfnamefont {Claudia}\
  \bibnamefont {Politi}}, \bibinfo {author} {\bibfnamefont {Lauritz}\
  \bibnamefont {Klaus}}, \bibinfo {author} {\bibfnamefont {Thomas}\
  \bibnamefont {Bland}}, \bibinfo {author} {\bibfnamefont {Manfred~J.}\
  \bibnamefont {Mark}}, \bibinfo {author} {\bibfnamefont {Luis}\ \bibnamefont
  {Santos}}, \bibinfo {author} {\bibfnamefont {Russell~N.}\ \bibnamefont
  {Bisset}}, \ and\ \bibinfo {author} {\bibfnamefont {Francesca}\ \bibnamefont
  {Ferlaino}},\ }\bibfield  {title} {\enquote {\bibinfo {title} {Can angular
  oscillations probe superfluidity in dipolar supersolids?}}\ }\href {\doibase
  10.1103/PhysRevLett.129.040403} {\bibfield  {journal} {\bibinfo  {journal}
  {Phys. Rev. Lett.}\ }\textbf {\bibinfo {volume} {129}},\ \bibinfo {pages}
  {040403} (\bibinfo {year} {2022})}\BibitemShut {NoStop}%
\bibitem [{\citenamefont {Bombín}\ \emph {et~al.}(2017)\citenamefont
  {Bombín}, \citenamefont {Boronat},\ and\ \citenamefont
  {Mazzanti}}]{Mazzanti:PRL:2017}%
  \BibitemOpen
  \bibfield  {author} {\bibinfo {author} {\bibfnamefont {R.}~\bibnamefont
  {Bombín}}, \bibinfo {author} {\bibfnamefont {J.}~\bibnamefont {Boronat}}, \
  and\ \bibinfo {author} {\bibfnamefont {F.}~\bibnamefont {Mazzanti}},\
  }\bibfield  {title} {\enquote {\bibinfo {title} {Dipolar bose supersolid
  stripes},}\ }\href {\doibase 10.1103/PhysRevLett.119.250402} {\bibfield
  {journal} {\bibinfo  {journal} {Phys. Rev. Lett.}\ }\textbf {\bibinfo
  {volume} {119}},\ \bibinfo {pages} {250402} (\bibinfo {year}
  {2017})}\BibitemShut {NoStop}%
\bibitem [{\citenamefont {Zhang}\ \emph {et~al.}(2019)\citenamefont {Zhang},
  \citenamefont {Maucher},\ and\ \citenamefont {Pohl}}]{zhang19}%
  \BibitemOpen
  \bibfield  {author} {\bibinfo {author} {\bibfnamefont {Yong-Chang}\
  \bibnamefont {Zhang}}, \bibinfo {author} {\bibfnamefont {Fabian}\
  \bibnamefont {Maucher}}, \ and\ \bibinfo {author} {\bibfnamefont {Thomas}\
  \bibnamefont {Pohl}},\ }\bibfield  {title} {\enquote {\bibinfo {title}
  {Supersolidity around a critical point in dipolar bose-einstein
  condensates},}\ }\href {\doibase 10.1103/PhysRevLett.123.015301} {\bibfield
  {journal} {\bibinfo  {journal} {Phys. Rev. Lett.}\ }\textbf {\bibinfo
  {volume} {123}},\ \bibinfo {pages} {015301} (\bibinfo {year}
  {2019})}\BibitemShut {NoStop}%
\bibitem [{\citenamefont {Blakie}\ \emph
  {et~al.}(2020{\natexlab{a}})\citenamefont {Blakie}, \citenamefont {Baillie},\
  and\ \citenamefont {Pal}}]{Blakie:CTP:2020}%
  \BibitemOpen
  \bibfield  {author} {\bibinfo {author} {\bibfnamefont {P~Blair}\ \bibnamefont
  {Blakie}}, \bibinfo {author} {\bibfnamefont {D}~\bibnamefont {Baillie}}, \
  and\ \bibinfo {author} {\bibfnamefont {Sukla}\ \bibnamefont {Pal}},\
  }\bibfield  {title} {\enquote {\bibinfo {title} {Variational theory for the
  ground state and collective excitations of an elongated dipolar
  condensate},}\ }\href {\doibase 10.1088/1572-9494/ab95fa} {\bibfield
  {journal} {\bibinfo  {journal} {Communications in Theoretical Physics}\
  }\textbf {\bibinfo {volume} {72}},\ \bibinfo {pages} {085501} (\bibinfo
  {year} {2020}{\natexlab{a}})}\BibitemShut {NoStop}%
\bibitem [{\citenamefont {Blakie}\ \emph
  {et~al.}(2020{\natexlab{b}})\citenamefont {Blakie}, \citenamefont {Baillie},
  \citenamefont {Chomaz},\ and\ \citenamefont {Ferlaino}}]{ferlaino20}%
  \BibitemOpen
  \bibfield  {author} {\bibinfo {author} {\bibfnamefont {P.~B.}\ \bibnamefont
  {Blakie}}, \bibinfo {author} {\bibfnamefont {D.}~\bibnamefont {Baillie}},
  \bibinfo {author} {\bibfnamefont {L.}~\bibnamefont {Chomaz}}, \ and\ \bibinfo
  {author} {\bibfnamefont {F.}~\bibnamefont {Ferlaino}},\ }\bibfield  {title}
  {\enquote {\bibinfo {title} {Supersolidity in an elongated dipolar
  condensate},}\ }\href {\doibase 10.1103/PhysRevResearch.2.043318} {\bibfield
  {journal} {\bibinfo  {journal} {Phys. Rev. Research}\ }\textbf {\bibinfo
  {volume} {2}},\ \bibinfo {pages} {043318} (\bibinfo {year}
  {2020}{\natexlab{b}})}\BibitemShut {NoStop}%
\bibitem [{\citenamefont {Zhang}\ \emph {et~al.}(2021)\citenamefont {Zhang},
  \citenamefont {Pohl},\ and\ \citenamefont {Maucher}}]{zhang21}%
  \BibitemOpen
  \bibfield  {author} {\bibinfo {author} {\bibfnamefont {Yong-Chang}\
  \bibnamefont {Zhang}}, \bibinfo {author} {\bibfnamefont {Thomas}\
  \bibnamefont {Pohl}}, \ and\ \bibinfo {author} {\bibfnamefont {Fabian}\
  \bibnamefont {Maucher}},\ }\bibfield  {title} {\enquote {\bibinfo {title}
  {Phases of supersolids in confined dipolar bose-einstein condensates},}\
  }\href {\doibase 10.1103/PhysRevA.104.013310} {\bibfield  {journal} {\bibinfo
   {journal} {Phys. Rev. A}\ }\textbf {\bibinfo {volume} {104}},\ \bibinfo
  {pages} {013310} (\bibinfo {year} {2021})}\BibitemShut {NoStop}%
\bibitem [{\citenamefont {Hertkorn}\ \emph
  {et~al.}(2021{\natexlab{a}})\citenamefont {Hertkorn}, \citenamefont
  {Schmidt}, \citenamefont {Guo}, \citenamefont {B\"ottcher}, \citenamefont
  {Ng}, \citenamefont {Graham}, \citenamefont {Uerlings}, \citenamefont
  {Langen}, \citenamefont {Zwierlein},\ and\ \citenamefont
  {Pfau}}]{Pfau:PRR:2021}%
  \BibitemOpen
  \bibfield  {author} {\bibinfo {author} {\bibfnamefont {J.}~\bibnamefont
  {Hertkorn}}, \bibinfo {author} {\bibfnamefont {J.-N.}\ \bibnamefont
  {Schmidt}}, \bibinfo {author} {\bibfnamefont {M.}~\bibnamefont {Guo}},
  \bibinfo {author} {\bibfnamefont {F.}~\bibnamefont {B\"ottcher}}, \bibinfo
  {author} {\bibfnamefont {K.~S.~H.}\ \bibnamefont {Ng}}, \bibinfo {author}
  {\bibfnamefont {S.~D.}\ \bibnamefont {Graham}}, \bibinfo {author}
  {\bibfnamefont {P.}~\bibnamefont {Uerlings}}, \bibinfo {author}
  {\bibfnamefont {T.}~\bibnamefont {Langen}}, \bibinfo {author} {\bibfnamefont
  {M.}~\bibnamefont {Zwierlein}}, \ and\ \bibinfo {author} {\bibfnamefont
  {T.}~\bibnamefont {Pfau}},\ }\bibfield  {title} {\enquote {\bibinfo {title}
  {Pattern formation in quantum ferrofluids: From supersolids to
  superglasses},}\ }\href {\doibase 10.1103/PhysRevResearch.3.033125}
  {\bibfield  {journal} {\bibinfo  {journal} {Phys. Rev. Res.}\ }\textbf
  {\bibinfo {volume} {3}},\ \bibinfo {pages} {033125} (\bibinfo {year}
  {2021}{\natexlab{a}})}\BibitemShut {NoStop}%
\bibitem [{\citenamefont {Hertkorn}\ \emph
  {et~al.}(2021{\natexlab{b}})\citenamefont {Hertkorn}, \citenamefont
  {Schmidt}, \citenamefont {Guo}, \citenamefont {B\"ottcher}, \citenamefont
  {Ng}, \citenamefont {Graham}, \citenamefont {Uerlings}, \citenamefont
  {B\"uchler}, \citenamefont {Langen}, \citenamefont {Zwierlein},\ and\
  \citenamefont {Pfau}}]{Pfau:PRL:2021}%
  \BibitemOpen
  \bibfield  {author} {\bibinfo {author} {\bibfnamefont {J.}~\bibnamefont
  {Hertkorn}}, \bibinfo {author} {\bibfnamefont {J.-N.}\ \bibnamefont
  {Schmidt}}, \bibinfo {author} {\bibfnamefont {M.}~\bibnamefont {Guo}},
  \bibinfo {author} {\bibfnamefont {F.}~\bibnamefont {B\"ottcher}}, \bibinfo
  {author} {\bibfnamefont {K.~S.~H.}\ \bibnamefont {Ng}}, \bibinfo {author}
  {\bibfnamefont {S.~D.}\ \bibnamefont {Graham}}, \bibinfo {author}
  {\bibfnamefont {P.}~\bibnamefont {Uerlings}}, \bibinfo {author}
  {\bibfnamefont {H.~P.}\ \bibnamefont {B\"uchler}}, \bibinfo {author}
  {\bibfnamefont {T.}~\bibnamefont {Langen}}, \bibinfo {author} {\bibfnamefont
  {M.}~\bibnamefont {Zwierlein}}, \ and\ \bibinfo {author} {\bibfnamefont
  {T.}~\bibnamefont {Pfau}},\ }\bibfield  {title} {\enquote {\bibinfo {title}
  {Supersolidity in two-dimensional trapped dipolar droplet arrays},}\ }\href
  {\doibase 10.1103/PhysRevLett.127.155301} {\bibfield  {journal} {\bibinfo
  {journal} {Phys. Rev. Lett.}\ }\textbf {\bibinfo {volume} {127}},\ \bibinfo
  {pages} {155301} (\bibinfo {year} {2021}{\natexlab{b}})}\BibitemShut
  {NoStop}%
\bibitem [{\citenamefont {W\"achtler}\ and\ \citenamefont
  {Santos}(2016)}]{wachtler16}%
  \BibitemOpen
  \bibfield  {author} {\bibinfo {author} {\bibfnamefont {F.}~\bibnamefont
  {W\"achtler}}\ and\ \bibinfo {author} {\bibfnamefont {L.}~\bibnamefont
  {Santos}},\ }\bibfield  {title} {\enquote {\bibinfo {title} {Quantum
  filaments in dipolar bose-einstein condensates},}\ }\href {\doibase
  10.1103/PhysRevA.93.061603} {\bibfield  {journal} {\bibinfo  {journal} {Phys.
  Rev. A}\ }\textbf {\bibinfo {volume} {93}},\ \bibinfo {pages} {061603}
  (\bibinfo {year} {2016})}\BibitemShut {NoStop}%
\bibitem [{\citenamefont {Bisset}\ \emph {et~al.}(2016)\citenamefont {Bisset},
  \citenamefont {Wilson}, \citenamefont {Baillie},\ and\ \citenamefont
  {Blakie}}]{Blakie:PRA:2016}%
  \BibitemOpen
  \bibfield  {author} {\bibinfo {author} {\bibfnamefont {R.~N.}\ \bibnamefont
  {Bisset}}, \bibinfo {author} {\bibfnamefont {R.~M.}\ \bibnamefont {Wilson}},
  \bibinfo {author} {\bibfnamefont {D.}~\bibnamefont {Baillie}}, \ and\
  \bibinfo {author} {\bibfnamefont {P.~B.}\ \bibnamefont {Blakie}},\ }\bibfield
   {title} {\enquote {\bibinfo {title} {Ground-state phase diagram of a dipolar
  condensate with quantum fluctuations},}\ }\href {\doibase
  10.1103/PhysRevA.94.033619} {\bibfield  {journal} {\bibinfo  {journal} {Phys.
  Rev. A}\ }\textbf {\bibinfo {volume} {94}},\ \bibinfo {pages} {033619}
  (\bibinfo {year} {2016})}\BibitemShut {NoStop}%
\bibitem [{\citenamefont {Baillie}\ \emph {et~al.}(2016)\citenamefont
  {Baillie}, \citenamefont {Wilson}, \citenamefont {Bisset},\ and\
  \citenamefont {Blakie}}]{Blakie:PRA2:2016}%
  \BibitemOpen
  \bibfield  {author} {\bibinfo {author} {\bibfnamefont {D.}~\bibnamefont
  {Baillie}}, \bibinfo {author} {\bibfnamefont {R.~M.}\ \bibnamefont {Wilson}},
  \bibinfo {author} {\bibfnamefont {R.~N.}\ \bibnamefont {Bisset}}, \ and\
  \bibinfo {author} {\bibfnamefont {P.~B.}\ \bibnamefont {Blakie}},\ }\bibfield
   {title} {\enquote {\bibinfo {title} {Self-bound dipolar droplet: A localized
  matter wave in free space},}\ }\href {\doibase 10.1103/PhysRevA.94.021602}
  {\bibfield  {journal} {\bibinfo  {journal} {Phys. Rev. A}\ }\textbf {\bibinfo
  {volume} {94}},\ \bibinfo {pages} {021602} (\bibinfo {year}
  {2016})}\BibitemShut {NoStop}%
\bibitem [{\citenamefont {B\"ottcher}\ \emph
  {et~al.}(2019{\natexlab{b}})\citenamefont {B\"ottcher}, \citenamefont
  {Wenzel}, \citenamefont {Schmidt}, \citenamefont {Guo}, \citenamefont
  {Langen}, \citenamefont {Ferrier-Barbut}, \citenamefont {Pfau}, \citenamefont
  {Bomb\'{\i}n}, \citenamefont {S\'anchez-Baena}, \citenamefont {Boronat},\
  and\ \citenamefont {Mazzanti}}]{bottcher19}%
  \BibitemOpen
  \bibfield  {author} {\bibinfo {author} {\bibfnamefont {Fabian}\ \bibnamefont
  {B\"ottcher}}, \bibinfo {author} {\bibfnamefont {Matthias}\ \bibnamefont
  {Wenzel}}, \bibinfo {author} {\bibfnamefont {Jan-Niklas}\ \bibnamefont
  {Schmidt}}, \bibinfo {author} {\bibfnamefont {Mingyang}\ \bibnamefont {Guo}},
  \bibinfo {author} {\bibfnamefont {Tim}\ \bibnamefont {Langen}}, \bibinfo
  {author} {\bibfnamefont {Igor}\ \bibnamefont {Ferrier-Barbut}}, \bibinfo
  {author} {\bibfnamefont {Tilman}\ \bibnamefont {Pfau}}, \bibinfo {author}
  {\bibfnamefont {Ra\'ul}\ \bibnamefont {Bomb\'{\i}n}}, \bibinfo {author}
  {\bibfnamefont {Joan}\ \bibnamefont {S\'anchez-Baena}}, \bibinfo {author}
  {\bibfnamefont {Jordi}\ \bibnamefont {Boronat}}, \ and\ \bibinfo {author}
  {\bibfnamefont {Ferran}\ \bibnamefont {Mazzanti}},\ }\bibfield  {title}
  {\enquote {\bibinfo {title} {Dilute dipolar quantum droplets beyond the
  extended gross-pitaevskii equation},}\ }\href {\doibase
  10.1103/PhysRevResearch.1.033088} {\bibfield  {journal} {\bibinfo  {journal}
  {Phys. Rev. Res.}\ }\textbf {\bibinfo {volume} {1}},\ \bibinfo {pages}
  {033088} (\bibinfo {year} {2019}{\natexlab{b}})}\BibitemShut {NoStop}%
\bibitem [{\citenamefont {Lima}\ and\ \citenamefont
  {Pelster}(2011)}]{Pelster:PRA:2011}%
  \BibitemOpen
  \bibfield  {author} {\bibinfo {author} {\bibfnamefont {Aristeu R.~P.}\
  \bibnamefont {Lima}}\ and\ \bibinfo {author} {\bibfnamefont {Axel}\
  \bibnamefont {Pelster}},\ }\bibfield  {title} {\enquote {\bibinfo {title}
  {Quantum fluctuations in dipolar bose gases},}\ }\href@noop {} {\bibfield
  {journal} {\bibinfo  {journal} {Phys. Rev. A}\ }\textbf {\bibinfo {volume}
  {84}},\ \bibinfo {pages} {041604(R)} (\bibinfo {year} {2011})}\BibitemShut
  {NoStop}%
\bibitem [{\citenamefont {Lima}\ and\ \citenamefont
  {Pelster}(2012)}]{pelster12}%
  \BibitemOpen
  \bibfield  {author} {\bibinfo {author} {\bibfnamefont {A.~R.~P.}\
  \bibnamefont {Lima}}\ and\ \bibinfo {author} {\bibfnamefont {A.}~\bibnamefont
  {Pelster}},\ }\bibfield  {title} {\enquote {\bibinfo {title} {Beyond
  mean-field low-lying excitations of dipolar bose gases},}\ }\href {\doibase
  10.1103/PhysRevA.86.063609} {\bibfield  {journal} {\bibinfo  {journal} {Phys.
  Rev. A}\ }\textbf {\bibinfo {volume} {86}},\ \bibinfo {pages} {063609}
  (\bibinfo {year} {2012})}\BibitemShut {NoStop}%
\bibitem [{\citenamefont {Kadau}\ \emph
  {et~al.}(2016{\natexlab{a}})\citenamefont {Kadau}, \citenamefont {Schmitt},
  \citenamefont {Wenzel}, \citenamefont {Wink}, \citenamefont {Maier},
  \citenamefont {Ferrier-Barbut},\ and\ \citenamefont
  {Pfau}}]{Pfau:nature:2016}%
  \BibitemOpen
  \bibfield  {author} {\bibinfo {author} {\bibfnamefont {Holger}\ \bibnamefont
  {Kadau}}, \bibinfo {author} {\bibfnamefont {Matthias}\ \bibnamefont
  {Schmitt}}, \bibinfo {author} {\bibfnamefont {Matthias}\ \bibnamefont
  {Wenzel}}, \bibinfo {author} {\bibfnamefont {Clarissa}\ \bibnamefont {Wink}},
  \bibinfo {author} {\bibfnamefont {Thomas}\ \bibnamefont {Maier}}, \bibinfo
  {author} {\bibfnamefont {Igor}\ \bibnamefont {Ferrier-Barbut}}, \ and\
  \bibinfo {author} {\bibfnamefont {Tilman}\ \bibnamefont {Pfau}},\ }\bibfield
  {title} {\enquote {\bibinfo {title} {Observing the rosensweig instability of
  a quantum ferrofluid},}\ }\href@noop {} {\bibfield  {journal} {\bibinfo
  {journal} {Nature}\ }\textbf {\bibinfo {volume} {530}},\ \bibinfo {pages}
  {194--197} (\bibinfo {year} {2016}{\natexlab{a}})}\BibitemShut {NoStop}%
\bibitem [{\citenamefont {Schmitt}\ \emph {et~al.}(2016)\citenamefont
  {Schmitt}, \citenamefont {Wenzel}, \citenamefont {B{\"o}ttcher},
  \citenamefont {Ferrier-Barbut},\ and\ \citenamefont
  {Pfau}}]{Pfau:nature2:2016}%
  \BibitemOpen
  \bibfield  {author} {\bibinfo {author} {\bibfnamefont {Matthias}\
  \bibnamefont {Schmitt}}, \bibinfo {author} {\bibfnamefont {Matthias}\
  \bibnamefont {Wenzel}}, \bibinfo {author} {\bibfnamefont {Fabian}\
  \bibnamefont {B{\"o}ttcher}}, \bibinfo {author} {\bibfnamefont {Igor}\
  \bibnamefont {Ferrier-Barbut}}, \ and\ \bibinfo {author} {\bibfnamefont
  {Tilman}\ \bibnamefont {Pfau}},\ }\bibfield  {title} {\enquote {\bibinfo
  {title} {Self-bound droplets of a dilute magnetic quantum liquid},}\
  }\href@noop {} {\bibfield  {journal} {\bibinfo  {journal} {Nature}\ }\textbf
  {\bibinfo {volume} {539}},\ \bibinfo {pages} {259--262} (\bibinfo {year}
  {2016})}\BibitemShut {NoStop}%
\bibitem [{\citenamefont {Saito}(2016)}]{Saito:JPhysJ:2016}%
  \BibitemOpen
  \bibfield  {author} {\bibinfo {author} {\bibfnamefont {Hiroki}\ \bibnamefont
  {Saito}},\ }\bibfield  {title} {\enquote {\bibinfo {title} {Path-integral
  monte carlo study on a droplet of a dipolar bose–einstein condensate
  stabilized by quantum fluctuation},}\ }\href@noop {} {\bibfield  {journal}
  {\bibinfo  {journal} {Journal of the Physical Society of Japan}\ }\textbf
  {\bibinfo {volume} {85}},\ \bibinfo {pages} {053001} (\bibinfo {year}
  {2016})}\BibitemShut {NoStop}%
\bibitem [{\citenamefont {Lahaye}\ \emph {et~al.}(2008)\citenamefont {Lahaye},
  \citenamefont {Metz}, \citenamefont {Fr\"ohlich}, \citenamefont {Koch},
  \citenamefont {Meister}, \citenamefont {Griesmaier}, \citenamefont {Pfau},
  \citenamefont {Saito}, \citenamefont {Kawaguchi},\ and\ \citenamefont
  {Ueda}}]{Lahaye:PRL:2008}%
  \BibitemOpen
  \bibfield  {author} {\bibinfo {author} {\bibfnamefont {T.}~\bibnamefont
  {Lahaye}}, \bibinfo {author} {\bibfnamefont {J.}~\bibnamefont {Metz}},
  \bibinfo {author} {\bibfnamefont {B.}~\bibnamefont {Fr\"ohlich}}, \bibinfo
  {author} {\bibfnamefont {T.}~\bibnamefont {Koch}}, \bibinfo {author}
  {\bibfnamefont {M.}~\bibnamefont {Meister}}, \bibinfo {author} {\bibfnamefont
  {A.}~\bibnamefont {Griesmaier}}, \bibinfo {author} {\bibfnamefont
  {T.}~\bibnamefont {Pfau}}, \bibinfo {author} {\bibfnamefont {H.}~\bibnamefont
  {Saito}}, \bibinfo {author} {\bibfnamefont {Y.}~\bibnamefont {Kawaguchi}}, \
  and\ \bibinfo {author} {\bibfnamefont {M.}~\bibnamefont {Ueda}},\ }\bibfield
  {title} {\enquote {\bibinfo {title} {$d$-wave collapse and explosion of a
  dipolar bose-einstein condensate},}\ }\href {\doibase
  10.1103/PhysRevLett.101.080401} {\bibfield  {journal} {\bibinfo  {journal}
  {Phys. Rev. Lett.}\ }\textbf {\bibinfo {volume} {101}},\ \bibinfo {pages}
  {080401} (\bibinfo {year} {2008})}\BibitemShut {NoStop}%
\bibitem [{\citenamefont {Koch}\ \emph {et~al.}(2008)\citenamefont {Koch},
  \citenamefont {Lahaye}, \citenamefont {Metz}, \citenamefont {Fr{\"o}hlich},
  \citenamefont {Griesmaier},\ and\ \citenamefont {Pfau}}]{Pfau:NatPhys:2008}%
  \BibitemOpen
  \bibfield  {author} {\bibinfo {author} {\bibfnamefont {T.}~\bibnamefont
  {Koch}}, \bibinfo {author} {\bibfnamefont {T.}~\bibnamefont {Lahaye}},
  \bibinfo {author} {\bibfnamefont {J.}~\bibnamefont {Metz}}, \bibinfo {author}
  {\bibfnamefont {B.}~\bibnamefont {Fr{\"o}hlich}}, \bibinfo {author}
  {\bibfnamefont {A.}~\bibnamefont {Griesmaier}}, \ and\ \bibinfo {author}
  {\bibfnamefont {T.}~\bibnamefont {Pfau}},\ }\bibfield  {title} {\enquote
  {\bibinfo {title} {Stabilization of a purely dipolar quantum gas against
  collapse},}\ }\href {\doibase 10.1038/nphys887} {\bibfield  {journal}
  {\bibinfo  {journal} {Nature Physics}\ }\textbf {\bibinfo {volume} {4}},\
  \bibinfo {pages} {218--222} (\bibinfo {year} {2008})}\BibitemShut {NoStop}%
\bibitem [{\citenamefont {Macia}\ \emph {et~al.}(2012)\citenamefont {Macia},
  \citenamefont {Hufnagl}, \citenamefont {Mazzanti}, \citenamefont {Boronat},\
  and\ \citenamefont {Zillich}}]{macia12}%
  \BibitemOpen
  \bibfield  {author} {\bibinfo {author} {\bibfnamefont {A.}~\bibnamefont
  {Macia}}, \bibinfo {author} {\bibfnamefont {D.}~\bibnamefont {Hufnagl}},
  \bibinfo {author} {\bibfnamefont {F.}~\bibnamefont {Mazzanti}}, \bibinfo
  {author} {\bibfnamefont {J.}~\bibnamefont {Boronat}}, \ and\ \bibinfo
  {author} {\bibfnamefont {R.~E.}\ \bibnamefont {Zillich}},\ }\bibfield
  {title} {\enquote {\bibinfo {title} {Excitations and stripe phase formation
  in a two-dimensional dipolar bose gas with tilted polarization},}\ }\href
  {\doibase 10.1103/PhysRevLett.109.235307} {\bibfield  {journal} {\bibinfo
  {journal} {Phys. Rev. Lett.}\ }\textbf {\bibinfo {volume} {109}},\ \bibinfo
  {pages} {235307} (\bibinfo {year} {2012})}\BibitemShut {NoStop}%
\bibitem [{\citenamefont {Macia}\ \emph {et~al.}(2014)\citenamefont {Macia},
  \citenamefont {Boronat},\ and\ \citenamefont {Mazzanti}}]{macia14}%
  \BibitemOpen
  \bibfield  {author} {\bibinfo {author} {\bibfnamefont {A.}~\bibnamefont
  {Macia}}, \bibinfo {author} {\bibfnamefont {J.}~\bibnamefont {Boronat}}, \
  and\ \bibinfo {author} {\bibfnamefont {F.}~\bibnamefont {Mazzanti}},\
  }\bibfield  {title} {\enquote {\bibinfo {title} {Phase diagram of dipolar
  bosons in two dimensions with tilted polarization},}\ }\href {\doibase
  10.1103/PhysRevA.90.061601} {\bibfield  {journal} {\bibinfo  {journal} {Phys.
  Rev. A}\ }\textbf {\bibinfo {volume} {90}},\ \bibinfo {pages} {061601}
  (\bibinfo {year} {2014})}\BibitemShut {NoStop}%
\bibitem [{\citenamefont {Gallem\'{\i}}\ and\ \citenamefont
  {Santos}(2022)}]{gallemi22}%
  \BibitemOpen
  \bibfield  {author} {\bibinfo {author} {\bibfnamefont {Albert}\ \bibnamefont
  {Gallem\'{\i}}}\ and\ \bibinfo {author} {\bibfnamefont {Luis}\ \bibnamefont
  {Santos}},\ }\bibfield  {title} {\enquote {\bibinfo {title} {Superfluid
  properties of a honeycomb dipolar supersolid},}\ }\href {\doibase
  10.1103/PhysRevA.106.063301} {\bibfield  {journal} {\bibinfo  {journal}
  {Phys. Rev. A}\ }\textbf {\bibinfo {volume} {106}},\ \bibinfo {pages}
  {063301} (\bibinfo {year} {2022})}\BibitemShut {NoStop}%
\bibitem [{\citenamefont {Guijarro}\ \emph {et~al.}(2022)\citenamefont
  {Guijarro}, \citenamefont {Astrakharchik},\ and\ \citenamefont
  {Boronat}}]{guijarro22}%
  \BibitemOpen
  \bibfield  {author} {\bibinfo {author} {\bibfnamefont {G.}~\bibnamefont
  {Guijarro}}, \bibinfo {author} {\bibfnamefont {G.~E.}\ \bibnamefont
  {Astrakharchik}}, \ and\ \bibinfo {author} {\bibfnamefont {J.}~\bibnamefont
  {Boronat}},\ }\bibfield  {title} {\enquote {\bibinfo {title} {Ultradilute
  quantum liquid of dipolar atoms in a bilayer},}\ }\href {\doibase
  10.1103/PhysRevLett.128.063401} {\bibfield  {journal} {\bibinfo  {journal}
  {Phys. Rev. Lett.}\ }\textbf {\bibinfo {volume} {128}},\ \bibinfo {pages}
  {063401} (\bibinfo {year} {2022})}\BibitemShut {NoStop}%
\bibitem [{\citenamefont {Staudinger}\ \emph {et~al.}(2023)\citenamefont
  {Staudinger}, \citenamefont {Hufnagl}, \citenamefont {Mazzanti},\ and\
  \citenamefont {Zillich}}]{mazzanti23}%
  \BibitemOpen
  \bibfield  {author} {\bibinfo {author} {\bibfnamefont {Clemens}\ \bibnamefont
  {Staudinger}}, \bibinfo {author} {\bibfnamefont {Diana}\ \bibnamefont
  {Hufnagl}}, \bibinfo {author} {\bibfnamefont {Ferran}\ \bibnamefont
  {Mazzanti}}, \ and\ \bibinfo {author} {\bibfnamefont {Robert~E.}\
  \bibnamefont {Zillich}},\ }\bibfield  {title} {\enquote {\bibinfo {title}
  {Striped dilute liquid of dipolar bosons in two dimensions},}\ }\href
  {\doibase 10.1103/PhysRevA.108.033303} {\bibfield  {journal} {\bibinfo
  {journal} {Phys. Rev. A}\ }\textbf {\bibinfo {volume} {108}},\ \bibinfo
  {pages} {033303} (\bibinfo {year} {2023})}\BibitemShut {NoStop}%
\bibitem [{\citenamefont {Aybar}\ and\ \citenamefont {Oktel}(2019)}]{oktel19}%
  \BibitemOpen
  \bibfield  {author} {\bibinfo {author} {\bibfnamefont {E.}~\bibnamefont
  {Aybar}}\ and\ \bibinfo {author} {\bibfnamefont {M.~\"O.}\ \bibnamefont
  {Oktel}},\ }\bibfield  {title} {\enquote {\bibinfo {title}
  {Temperature-dependent density profiles of dipolar droplets},}\ }\href
  {\doibase 10.1103/PhysRevA.99.013620} {\bibfield  {journal} {\bibinfo
  {journal} {Phys. Rev. A}\ }\textbf {\bibinfo {volume} {99}},\ \bibinfo
  {pages} {013620} (\bibinfo {year} {2019})}\BibitemShut {NoStop}%
\bibitem [{\citenamefont {\"Ozt\"urk}\ \emph {et~al.}(2020)\citenamefont
  {\"Ozt\"urk}, \citenamefont {Aybar},\ and\ \citenamefont {Oktel}}]{oktel20}%
  \BibitemOpen
  \bibfield  {author} {\bibinfo {author} {\bibfnamefont {S.~F.}\ \bibnamefont
  {\"Ozt\"urk}}, \bibinfo {author} {\bibfnamefont {Enes}\ \bibnamefont
  {Aybar}}, \ and\ \bibinfo {author} {\bibfnamefont {M.~\"O.}\ \bibnamefont
  {Oktel}},\ }\bibfield  {title} {\enquote {\bibinfo {title} {Temperature
  dependence of the density and excitations of dipolar droplets},}\ }\href
  {\doibase 10.1103/PhysRevA.102.033329} {\bibfield  {journal} {\bibinfo
  {journal} {Phys. Rev. A}\ }\textbf {\bibinfo {volume} {102}},\ \bibinfo
  {pages} {033329} (\bibinfo {year} {2020})}\BibitemShut {NoStop}%
\bibitem [{\citenamefont {Sohmen}\ \emph
  {et~al.}(2021{\natexlab{a}})\citenamefont {Sohmen}, \citenamefont {Politi},
  \citenamefont {Klaus}, \citenamefont {Chomaz}, \citenamefont {Mark},
  \citenamefont {Norcia},\ and\ \citenamefont {Ferlaino}}]{Ferlaino:PRL:2021}%
  \BibitemOpen
  \bibfield  {author} {\bibinfo {author} {\bibfnamefont {Maximilian}\
  \bibnamefont {Sohmen}}, \bibinfo {author} {\bibfnamefont {Claudia}\
  \bibnamefont {Politi}}, \bibinfo {author} {\bibfnamefont {Lauritz}\
  \bibnamefont {Klaus}}, \bibinfo {author} {\bibfnamefont {Lauriane}\
  \bibnamefont {Chomaz}}, \bibinfo {author} {\bibfnamefont {Manfred~J.}\
  \bibnamefont {Mark}}, \bibinfo {author} {\bibfnamefont {Matthew~A.}\
  \bibnamefont {Norcia}}, \ and\ \bibinfo {author} {\bibfnamefont {Francesca}\
  \bibnamefont {Ferlaino}},\ }\bibfield  {title} {\enquote {\bibinfo {title}
  {Birth, life, and death of a dipolar supersolid},}\ }\href {\doibase
  10.1103/PhysRevLett.126.233401} {\bibfield  {journal} {\bibinfo  {journal}
  {Phys. Rev. Lett.}\ }\textbf {\bibinfo {volume} {126}},\ \bibinfo {pages}
  {233401} (\bibinfo {year} {2021}{\natexlab{a}})}\BibitemShut {NoStop}%
\bibitem [{\citenamefont {S{\'a}nchez-Baena}\ \emph {et~al.}(2023)\citenamefont
  {S{\'a}nchez-Baena}, \citenamefont {Politi}, \citenamefont {Maucher},
  \citenamefont {Ferlaino},\ and\ \citenamefont {Pohl}}]{baena22}%
  \BibitemOpen
  \bibfield  {author} {\bibinfo {author} {\bibfnamefont {J.}~\bibnamefont
  {S{\'a}nchez-Baena}}, \bibinfo {author} {\bibfnamefont {C.}~\bibnamefont
  {Politi}}, \bibinfo {author} {\bibfnamefont {F.}~\bibnamefont {Maucher}},
  \bibinfo {author} {\bibfnamefont {F.}~\bibnamefont {Ferlaino}}, \ and\
  \bibinfo {author} {\bibfnamefont {T.}~\bibnamefont {Pohl}},\ }\bibfield
  {title} {\enquote {\bibinfo {title} {Heating a dipolar quantum fluid into a
  solid},}\ }\href {\doibase 10.1038/s41467-023-37207-3} {\bibfield  {journal}
  {\bibinfo  {journal} {Nature Communications}\ }\textbf {\bibinfo {volume}
  {14}},\ \bibinfo {pages} {1868} (\bibinfo {year} {2023})}\BibitemShut
  {NoStop}%
\bibitem [{\citenamefont {Smith}\ \emph {et~al.}(2022)\citenamefont {Smith},
  \citenamefont {Baillie},\ and\ \citenamefont {Blakie}}]{smith22}%
  \BibitemOpen
  \bibfield  {author} {\bibinfo {author} {\bibfnamefont {Joseph~C.}\
  \bibnamefont {Smith}}, \bibinfo {author} {\bibfnamefont {D.}~\bibnamefont
  {Baillie}}, \ and\ \bibinfo {author} {\bibfnamefont {P.~B.}\ \bibnamefont
  {Blakie}},\ }\bibfield  {title} {\enquote {\bibinfo {title} {Supersolidity
  and crystallization of a dipolar bose gas in an infinite tube},}\ }\href
  {\doibase 10.48550/ARXIV.2212.07607} {\  (\bibinfo {year} {2022}),\
  10.48550/ARXIV.2212.07607}\BibitemShut {NoStop}%
\bibitem [{\citenamefont {Ilg}\ and\ \citenamefont
  {B\"uchler}(2023)}]{Buchler:PRA:2022}%
  \BibitemOpen
  \bibfield  {author} {\bibinfo {author} {\bibfnamefont {Tobias}\ \bibnamefont
  {Ilg}}\ and\ \bibinfo {author} {\bibfnamefont {Hans~Peter}\ \bibnamefont
  {B\"uchler}},\ }\bibfield  {title} {\enquote {\bibinfo {title} {Ground-state
  stability and excitation spectrum of a one-dimensional dipolar supersolid},}\
  }\href {\doibase 10.1103/PhysRevA.107.013314} {\bibfield  {journal} {\bibinfo
   {journal} {Phys. Rev. A}\ }\textbf {\bibinfo {volume} {107}},\ \bibinfo
  {pages} {013314} (\bibinfo {year} {2023})}\BibitemShut {NoStop}%
\bibitem [{\citenamefont {Giorgini}\ \emph {et~al.}(1997)\citenamefont
  {Giorgini}, \citenamefont {Pitaevskii},\ and\ \citenamefont
  {Stringari}}]{Giorgini1997}%
  \BibitemOpen
  \bibfield  {author} {\bibinfo {author} {\bibfnamefont {S.}~\bibnamefont
  {Giorgini}}, \bibinfo {author} {\bibfnamefont {L.~P.}\ \bibnamefont
  {Pitaevskii}}, \ and\ \bibinfo {author} {\bibfnamefont {S.}~\bibnamefont
  {Stringari}},\ }\bibfield  {title} {\enquote {\bibinfo {title}
  {Thermodynamics of a trapped bose-condensed gas},}\ }\href {\doibase
  10.1007/BF02396737} {\bibfield  {journal} {\bibinfo  {journal} {Journal of
  Low Temperature Physics}\ }\textbf {\bibinfo {volume} {109}},\ \bibinfo
  {pages} {309--355} (\bibinfo {year} {1997})}\BibitemShut {NoStop}%
\bibitem [{\citenamefont {Griffin}(1996)}]{Griffin:PRB:1996}%
  \BibitemOpen
  \bibfield  {author} {\bibinfo {author} {\bibfnamefont {A.}~\bibnamefont
  {Griffin}},\ }\bibfield  {title} {\enquote {\bibinfo {title} {Conserving and
  gapless approximations for an inhomogeneous bose gas at finite
  temperatures},}\ }\href {\doibase 10.1103/PhysRevB.53.9341} {\bibfield
  {journal} {\bibinfo  {journal} {Phys. Rev. B}\ }\textbf {\bibinfo {volume}
  {53}},\ \bibinfo {pages} {9341--9347} (\bibinfo {year} {1996})}\BibitemShut
  {NoStop}%
\bibitem [{\citenamefont {Shi}\ and\ \citenamefont {Griffin}(1998)}]{shi98}%
  \BibitemOpen
  \bibfield  {author} {\bibinfo {author} {\bibfnamefont {Hua}\ \bibnamefont
  {Shi}}\ and\ \bibinfo {author} {\bibfnamefont {Allan}\ \bibnamefont
  {Griffin}},\ }\bibfield  {title} {\enquote {\bibinfo {title}
  {Finite-temperature excitations in a dilute bose-condensed gas},}\ }\href
  {\doibase https://doi.org/10.1016/S0370-1573(98)00015-5} {\bibfield
  {journal} {\bibinfo  {journal} {Physics Reports}\ }\textbf {\bibinfo {volume}
  {304}},\ \bibinfo {pages} {1--87} (\bibinfo {year} {1998})}\BibitemShut
  {NoStop}%
\bibitem [{\citenamefont {Cormack}\ and\ \citenamefont
  {Hutchinson}(2012)}]{cormack12}%
  \BibitemOpen
  \bibfield  {author} {\bibinfo {author} {\bibfnamefont {S.~C.}\ \bibnamefont
  {Cormack}}\ and\ \bibinfo {author} {\bibfnamefont {D.~A.~W.}\ \bibnamefont
  {Hutchinson}},\ }\bibfield  {title} {\enquote {\bibinfo {title}
  {Finite-temperature dipolar ultracold bose gas with exchange interactions},}\
  }\href {\doibase 10.1103/PhysRevA.86.053619} {\bibfield  {journal} {\bibinfo
  {journal} {Phys. Rev. A}\ }\textbf {\bibinfo {volume} {86}},\ \bibinfo
  {pages} {053619} (\bibinfo {year} {2012})}\BibitemShut {NoStop}%
\bibitem [{\citenamefont {Wang}\ \emph {et~al.}(2021)\citenamefont {Wang},
  \citenamefont {Liu},\ and\ \citenamefont {Hu}}]{wang21}%
  \BibitemOpen
  \bibfield  {author} {\bibinfo {author} {\bibfnamefont {Jia}\ \bibnamefont
  {Wang}}, \bibinfo {author} {\bibfnamefont {Xia-Ji}\ \bibnamefont {Liu}}, \
  and\ \bibinfo {author} {\bibfnamefont {Hui}\ \bibnamefont {Hu}},\ }\bibfield
  {title} {\enquote {\bibinfo {title} {Ultradilute self-bound quantum droplets
  in bose–bose mixtures at finite temperature*},}\ }\href {\doibase
  10.1088/1674-1056/abd2ad} {\bibfield  {journal} {\bibinfo  {journal} {Chinese
  Physics B}\ }\textbf {\bibinfo {volume} {30}},\ \bibinfo {pages} {010306}
  (\bibinfo {year} {2021})}\BibitemShut {NoStop}%
\bibitem [{\citenamefont {Kadau}\ \emph
  {et~al.}(2016{\natexlab{b}})\citenamefont {Kadau}, \citenamefont {Schmitt},
  \citenamefont {Wenzel}, \citenamefont {Wink}, \citenamefont {Maier},
  \citenamefont {Ferrier-Barbut},\ and\ \citenamefont {Pfau}}]{kadau16}%
  \BibitemOpen
  \bibfield  {author} {\bibinfo {author} {\bibfnamefont {Holger}\ \bibnamefont
  {Kadau}}, \bibinfo {author} {\bibfnamefont {Matthias}\ \bibnamefont
  {Schmitt}}, \bibinfo {author} {\bibfnamefont {Matthias}\ \bibnamefont
  {Wenzel}}, \bibinfo {author} {\bibfnamefont {Clarissa}\ \bibnamefont {Wink}},
  \bibinfo {author} {\bibfnamefont {Thomas}\ \bibnamefont {Maier}}, \bibinfo
  {author} {\bibfnamefont {Igor}\ \bibnamefont {Ferrier-Barbut}}, \ and\
  \bibinfo {author} {\bibfnamefont {Tilman}\ \bibnamefont {Pfau}},\ }\bibfield
  {title} {\enquote {\bibinfo {title} {Observing the rosensweig instability of
  a quantum ferrofluid},}\ }\href {\doibase 10.1038/nature16485} {\bibfield
  {journal} {\bibinfo  {journal} {Nature}\ }\textbf {\bibinfo {volume} {530}},\
  \bibinfo {pages} {194--197} (\bibinfo {year}
  {2016}{\natexlab{b}})}\BibitemShut {NoStop}%
\bibitem [{\citenamefont {Chomaz}\ \emph {et~al.}(2016)\citenamefont {Chomaz},
  \citenamefont {Baier}, \citenamefont {Petter}, \citenamefont {Mark},
  \citenamefont {W\"achtler}, \citenamefont {Santos},\ and\ \citenamefont
  {Ferlaino}}]{ferlaino16}%
  \BibitemOpen
  \bibfield  {author} {\bibinfo {author} {\bibfnamefont {L.}~\bibnamefont
  {Chomaz}}, \bibinfo {author} {\bibfnamefont {S.}~\bibnamefont {Baier}},
  \bibinfo {author} {\bibfnamefont {D.}~\bibnamefont {Petter}}, \bibinfo
  {author} {\bibfnamefont {M.~J.}\ \bibnamefont {Mark}}, \bibinfo {author}
  {\bibfnamefont {F.}~\bibnamefont {W\"achtler}}, \bibinfo {author}
  {\bibfnamefont {L.}~\bibnamefont {Santos}}, \ and\ \bibinfo {author}
  {\bibfnamefont {F.}~\bibnamefont {Ferlaino}},\ }\bibfield  {title} {\enquote
  {\bibinfo {title} {Quantum-fluctuation-driven crossover from a dilute
  bose-einstein condensate to a macrodroplet in a dipolar quantum fluid},}\
  }\href {\doibase 10.1103/PhysRevX.6.041039} {\bibfield  {journal} {\bibinfo
  {journal} {Phys. Rev. X}\ }\textbf {\bibinfo {volume} {6}},\ \bibinfo {pages}
  {041039} (\bibinfo {year} {2016})}\BibitemShut {NoStop}%
\bibitem [{\citenamefont {Leggett}(1970{\natexlab{b}})}]{legget70}%
  \BibitemOpen
  \bibfield  {author} {\bibinfo {author} {\bibfnamefont {A.~J.}\ \bibnamefont
  {Leggett}},\ }\bibfield  {title} {\enquote {\bibinfo {title} {Can a solid be
  "superfluid"?}}\ }\href {\doibase 10.1103/PhysRevLett.25.1543} {\bibfield
  {journal} {\bibinfo  {journal} {Phys. Rev. Lett.}\ }\textbf {\bibinfo
  {volume} {25}},\ \bibinfo {pages} {1543--1546} (\bibinfo {year}
  {1970}{\natexlab{b}})}\BibitemShut {NoStop}%
\bibitem [{\citenamefont {Sep\'ulveda}\ \emph {et~al.}(2008)\citenamefont
  {Sep\'ulveda}, \citenamefont {Josserand},\ and\ \citenamefont
  {Rica}}]{sepulveda08}%
  \BibitemOpen
  \bibfield  {author} {\bibinfo {author} {\bibfnamefont {N\'estor}\
  \bibnamefont {Sep\'ulveda}}, \bibinfo {author} {\bibfnamefont {Christophe}\
  \bibnamefont {Josserand}}, \ and\ \bibinfo {author} {\bibfnamefont {Sergio}\
  \bibnamefont {Rica}},\ }\bibfield  {title} {\enquote {\bibinfo {title}
  {Nonclassical rotational inertia fraction in a one-dimensional model of a
  supersolid},}\ }\href {\doibase 10.1103/PhysRevB.77.054513} {\bibfield
  {journal} {\bibinfo  {journal} {Phys. Rev. B}\ }\textbf {\bibinfo {volume}
  {77}},\ \bibinfo {pages} {054513} (\bibinfo {year} {2008})}\BibitemShut
  {NoStop}%
\bibitem [{\citenamefont {Leggett}(1998)}]{leggett98}%
  \BibitemOpen
  \bibfield  {author} {\bibinfo {author} {\bibfnamefont {A.~J.}\ \bibnamefont
  {Leggett}},\ }\bibfield  {title} {\enquote {\bibinfo {title} {On the
  superfluid fraction of an arbitrary many-body system at t=0},}\ }\href
  {\doibase 10.1023/B:JOSS.0000033170.38619.6c} {\bibfield  {journal} {\bibinfo
   {journal} {Journal of Statistical Physics}\ }\textbf {\bibinfo {volume}
  {93}},\ \bibinfo {pages} {927--941} (\bibinfo {year} {1998})}\BibitemShut
  {NoStop}%
\bibitem [{\citenamefont {Sohmen}\ \emph
  {et~al.}(2021{\natexlab{b}})\citenamefont {Sohmen}, \citenamefont {Politi},
  \citenamefont {Klaus}, \citenamefont {Chomaz}, \citenamefont {Mark},
  \citenamefont {Norcia},\ and\ \citenamefont {Ferlaino}}]{ferlaino21}%
  \BibitemOpen
  \bibfield  {author} {\bibinfo {author} {\bibfnamefont {Maximilian}\
  \bibnamefont {Sohmen}}, \bibinfo {author} {\bibfnamefont {Claudia}\
  \bibnamefont {Politi}}, \bibinfo {author} {\bibfnamefont {Lauritz}\
  \bibnamefont {Klaus}}, \bibinfo {author} {\bibfnamefont {Lauriane}\
  \bibnamefont {Chomaz}}, \bibinfo {author} {\bibfnamefont {Manfred~J.}\
  \bibnamefont {Mark}}, \bibinfo {author} {\bibfnamefont {Matthew~A.}\
  \bibnamefont {Norcia}}, \ and\ \bibinfo {author} {\bibfnamefont {Francesca}\
  \bibnamefont {Ferlaino}},\ }\bibfield  {title} {\enquote {\bibinfo {title}
  {Birth, life, and death of a dipolar supersolid},}\ }\href {\doibase
  10.1103/PhysRevLett.126.233401} {\bibfield  {journal} {\bibinfo  {journal}
  {Phys. Rev. Lett.}\ }\textbf {\bibinfo {volume} {126}},\ \bibinfo {pages}
  {233401} (\bibinfo {year} {2021}{\natexlab{b}})}\BibitemShut {NoStop}%
\bibitem [{\citenamefont {Hertkorn}\ \emph
  {et~al.}(2021{\natexlab{c}})\citenamefont {Hertkorn}, \citenamefont
  {Schmidt}, \citenamefont {Guo}, \citenamefont {B\"ottcher}, \citenamefont
  {Ng}, \citenamefont {Graham}, \citenamefont {Uerlings}, \citenamefont
  {Langen}, \citenamefont {Zwierlein},\ and\ \citenamefont
  {Pfau}}]{hertkorn21}%
  \BibitemOpen
  \bibfield  {author} {\bibinfo {author} {\bibfnamefont {J.}~\bibnamefont
  {Hertkorn}}, \bibinfo {author} {\bibfnamefont {J.-N.}\ \bibnamefont
  {Schmidt}}, \bibinfo {author} {\bibfnamefont {M.}~\bibnamefont {Guo}},
  \bibinfo {author} {\bibfnamefont {F.}~\bibnamefont {B\"ottcher}}, \bibinfo
  {author} {\bibfnamefont {K.~S.~H.}\ \bibnamefont {Ng}}, \bibinfo {author}
  {\bibfnamefont {S.~D.}\ \bibnamefont {Graham}}, \bibinfo {author}
  {\bibfnamefont {P.}~\bibnamefont {Uerlings}}, \bibinfo {author}
  {\bibfnamefont {T.}~\bibnamefont {Langen}}, \bibinfo {author} {\bibfnamefont
  {M.}~\bibnamefont {Zwierlein}}, \ and\ \bibinfo {author} {\bibfnamefont
  {T.}~\bibnamefont {Pfau}},\ }\bibfield  {title} {\enquote {\bibinfo {title}
  {Pattern formation in quantum ferrofluids: From supersolids to
  superglasses},}\ }\href {\doibase 10.1103/PhysRevResearch.3.033125}
  {\bibfield  {journal} {\bibinfo  {journal} {Phys. Rev. Res.}\ }\textbf
  {\bibinfo {volume} {3}},\ \bibinfo {pages} {033125} (\bibinfo {year}
  {2021}{\natexlab{c}})}\BibitemShut {NoStop}%
\bibitem [{\citenamefont {Norcia}\ \emph {et~al.}(2021)\citenamefont {Norcia},
  \citenamefont {Politi}, \citenamefont {Klaus}, \citenamefont {Poli},
  \citenamefont {Sohmen}, \citenamefont {Mark}, \citenamefont {Bisset},
  \citenamefont {Santos},\ and\ \citenamefont {Ferlaino}}]{norcia21}%
  \BibitemOpen
  \bibfield  {author} {\bibinfo {author} {\bibfnamefont {Matthew~A.}\
  \bibnamefont {Norcia}}, \bibinfo {author} {\bibfnamefont {Claudia}\
  \bibnamefont {Politi}}, \bibinfo {author} {\bibfnamefont {Lauritz}\
  \bibnamefont {Klaus}}, \bibinfo {author} {\bibfnamefont {Elena}\ \bibnamefont
  {Poli}}, \bibinfo {author} {\bibfnamefont {Maximilian}\ \bibnamefont
  {Sohmen}}, \bibinfo {author} {\bibfnamefont {Manfred~J.}\ \bibnamefont
  {Mark}}, \bibinfo {author} {\bibfnamefont {Russell~N.}\ \bibnamefont
  {Bisset}}, \bibinfo {author} {\bibfnamefont {Luis}\ \bibnamefont {Santos}}, \
  and\ \bibinfo {author} {\bibfnamefont {Francesca}\ \bibnamefont {Ferlaino}},\
  }\bibfield  {title} {\enquote {\bibinfo {title} {Two-dimensional
  supersolidity in a dipolar quantum gas},}\ }\href {\doibase
  10.1038/s41586-021-03725-7} {\bibfield  {journal} {\bibinfo  {journal}
  {Nature}\ }\textbf {\bibinfo {volume} {596}},\ \bibinfo {pages} {357--361}
  (\bibinfo {year} {2021})}\BibitemShut {NoStop}%
\end{thebibliography}%

\widetext

\end{document}